\documentclass[12pt,preprint]{aastex} 

\usepackage{emulateapj5}
\usepackage{epsfig} 
\usepackage{apjfonts}

\newcommand{\mdot}{M$_{\odot}$}
\newcommand{\mv}{$M_V$}

\newcommand{\opt}{$\mathrm{_{opt}}$}
\newcommand{\hst}{{\it HST}}
\newcommand{\cha}{{\it Chandra}}

\newcommand{\lx}{$L_{x}$}
\newcommand{\fx}{$F_{x}$}
\newcommand{\hal}{H$\alpha$}
\newcommand{\fxfopt}{$F_X/F_{opt}$}
\newcommand{\ergs}{erg~s$^{-1}$}

\shorttitle{Optical Counterparts to 47~Tuc X-ray sources. II}
\shortauthors{Edmonds et al.}

\begin{document}

\title{An Extensive Census of \hst\ Counterparts to \cha\ X-ray Sources in
the Globular Cluster 47~Tucanae. II. Time Series and
Analysis\altaffilmark{1}}

\altaffiltext{1}{Based on observations with the NASA/ESA {\it Hubble Space
Telescope} obtained at STScI, which is operated by AURA, Inc. under NASA
contract NAS 5-26555.}

\author{Peter D. Edmonds\altaffilmark{2}, Ronald
L. Gilliland\altaffilmark{3}, Craig O. Heinke\altaffilmark{2}, \& Jonathan
E. Grindlay\altaffilmark{2}} \altaffiltext{2}{Harvard-Smithsonian Center
for Astrophysics, 60 Garden St, Cambridge, MA 02138;
pedmonds@cfa.harvard.edu; cheinke@cfa.harvard.edu, josh@cfa.harvard.edu}
\altaffiltext{3}{Space Telescope Science Institute, 3700 San Martin Drive,
Baltimore, MD 21218; gillil@stsci.edu}

\begin{abstract}

We report time series and variability information for the optical
identifications of X-ray sources in 47~Tucanae reported in Paper~I (at
least 22 CVs and 29 active binaries).  The radial distribution of the CVs
is indistinguishable from that of the MSPs detected by Freire et
al. (2001). A study of the eight CVs with secure orbital periods (two
obtained from the \cha\ study of Grindlay et al. 2001a) shows that the
47~Tuc CVs have fainter accretion disks, in the $V$ band, than field CVs
with similar periods.  These faint disks and the faint absolute magnitudes
(\mv) of the 47~Tuc CVs suggests they have low accretion rates. One
possible explanation is that the 47~Tuc objects may be a more
representative sample of CVs, down to our detection threshold, than the CVs
found in the field (where many low accretion rate systems are believed to
be undiscovered), showing the advantages of deep globular cluster
observations. The median \fxfopt\ value for the 47~Tuc CVs is higher than
that of all known classes of field CV, partly because of the faint \mv\
values and partly because of the relatively high X-ray luminosities (\lx).
The latter are only seen in DQ~Her systems in the field, but the 47~Tuc CVs
are much fainter optically than most field DQ~Hers.  Previous work by
Edmonds et al. (1999) has shown that the 4 brightest CVs in NGC~6397 have
optical spectra and broadband colors that are consistent with DQ~Hers
having lower than average accretion rates.  Some combination of magnetic
behavior and low accretion rates may be able to explain our observations,
but the results at present are ambiguous, since no class of field CV has
distributions of both \lx\ and \mv\ that are consistent with those of the
47~Tuc CVs.

The radial distribution of the X-ray detected active binaries is
indistinguishable from that of the much larger sample of optical variables
(eclipsing and contact binaries and BY Dra variables) detected in previous
WFPC2 studies by Albrow et al (2001). The X-ray properties of these objects
(luminosity, hardness ratios and variability) are consistent with those of
active binaries found in field studies, and the \fxfopt\ distribution is
significantly different from those of the CVs and the millisecond pulsars
(MSPs) that are detected (or possibly detected) in the optical.  Despite
these results, we examine the possibility that a few of the active binaries
are MSPs with main sequence companions resulting from double exchanges in
the crowded core of 47~Tuc. No solid evidence is found for a significant
population of such objects, and therefore using the methods of Grindlay et
al. (2002) we estimate that the number of MSPs in 47~Tuc with luminosities
$>10^{30}$\ergs\ is $\sim$30--40, near the previous lower limit.

We present the results of a new, deeper search for faint low-mass X-ray
binaries (LMXBs) in quiescence.  One reasonable and one marginal candidate
for optical identification of a quiescent LMXB was found (one is already
known).  Finally, it is shown that the periods of the blue variables
showing little or no evidence for X-ray emission are too long for Roche
lobe filling (if the variations are ellipsoidal). These blue variables also
show no evidence for the large flickering levels seen in comparably bright
CVs. At present we have no satisfactory explanation for these objects,
although some may be detached WD-main sequence star binaries.

\end{abstract}

\keywords{binaries: general -- globular clusters: individual (47~Tucanae)
-- techniques: photometric -- X-rays: binaries}

\section{Introduction}
\label{sect.intro}

Binaries are well known to have a profound impact on the dynamical
evolution of globular clusters (Hut et al. 1992 and Edmonds et al. 2003,
hereafter Paper~I), and they offer an opportunity to study the results of
stellar interactions. They also offer the crucial advantages of studying
binary systems at the same, well-determined distance, age, metallicity and
reddening. Accurate distances are particularly important for reducing
uncertainties in luminosities, and making reliable cluster-to-cluster
comparisons and cluster-to-field comparisons.

Despite these promising goals, observational progress in detecting globular
cluster binaries was initially slow, mainly because of spatial resolution
and sensitivity limitations. However, the superb imaging capabilities of
\cha\ and \hst\ has now resulted in the detection of large numbers of
compact binaries in several globular clusters (see Grindlay et al. 2001a,
hereafter GHE01a and references in Paper~I).  In particular, using the
\cha\ data of GHE01a and the \hst\ data of Gilliland et al. (2000; \hst\
program GO-8267), a large population of X-ray binaries with optical
identifications has been found in 47~Tuc, with over 50 binaries reported in
Paper~I. There, the details of the astrometry and photometry for this
sample were reported, with most of the systems being identified as either
accreting white dwarf-main sequence star binaries (cataclysmic variables or
CVs) or chromospherically {\it active binaries}.

This paper reports \hst\ optical time series and other detailed analysis
for the sample of binaries reported in Paper~I. Variability for most of the
CVs is detected, confirming the CV identifications given in Paper~I based
on absolute photometry alone. In particular flickering for the brighter
optical IDs is found, plus periods for several of the higher inclination
systems and some long-term variability. The CV periods and the absolute
magnitudes determined in Paper~I are compared with predictions for
Roche-lobe filling secondaries and are compared with CVs found in the
field. Time series for the active binaries are also presented here, with an
emphasis on the variables not discussed by Albrow et al. (2001; hereafter
AGB01).

The radial distribution (radial offsets from cluster center) of the CVs and
active binaries discovered in Paper~I are compared with the radial
distribution of MSPs from Freire et al. (2001) and with the total stellar
population, to test whether these distributions are consistent with the
expectations of mass segregation. Comparisons between the radial
distributions of the X-ray detected active binaries and the larger sample
of binaries discovered by AGB01 are also made.

Other important diagnostics examined here are the X-ray luminosities and
optical magnitudes. The flux ratio between the two was shown by Richman
(1996) and Verbunt et al. (1997; hereafter VBR97) to be inversely
proportional to the CV accretion rate (for non-magnetic systems), and here
we make detailed comparisons between the 47~Tuc CVs and field CVs to help
constrain the CV accretion rates in 47~Tuc. The X-ray to optical flux ratio
is also useful in searching for additional quiescent low-mass X-ray
binaries (qLMXBs) besides X5 and X7 (GHE01a and Heinke et al. 2003). We
also discuss the X-ray luminosity and absolute magnitude distributions of
the 47~Tuc CVs and compare to field systems.

The time series will be described in \S~\ref{sect.tseries}, followed by an
analysis section (\S~\ref{sect.anal}), including a study of the radial
distribution of the sources and their X-ray to optical flux ratios. This
will be followed by the discussion in \S~\ref{sect.disc}, and an
interpretation of the results given in both Papers~I and II.

\section{Time Series}
\label{sect.tseries}

\subsection{Optical data}

Paper~I discusses the photometric results for the optical IDs (including
the CMDs) in detail.  Exquisite time series were produced in the $V$ and
$I$ bands for the GO-8267 dataset, and are analyzed in this paper. Detailed
simulations were carried out by AGB01, giving the recovery rate of
simulated variables as a function of period, $V$-magnitude and signal
amplitude for the PC1 and WF2 chips. For amplitudes of
$\Delta$(intensity)/intensity = 0.1, the recovery rate for PC1 was
$\sim$100\% from the MSTO to $V$=23.5 for all input periods below 3.2 days,
and for amplitudes of 0.01 the recovery rate fell from $\sim$70\% at $V$=20
to $<$10\% at $V$=21.5 (AGB01). The WF2 recovery rates extend to slightly
fainter magnitudes (AGB01), but the generally higher crowding and
saturation levels in the WF2 images counteract this gain, for variable
searches.

\subsection{Cataclysmic variables}
\label{sect.cvtseries}

\subsubsection{GO-8267 data}

For the CV candidates found in the GO-8267 FoV (Paper~I) we have analyzed
the high quality time series produced by the planet and binary searches of
Gilliland et al. (2000) and AGB01. Time series were produced using
difference image analysis combined with aperture and PSF fitting photometry
(see Gilliland et al. 2000 and AGB01 for more details). The size of the
region used in the photometry extractions was decreased to as small as one
pixel in the cases where plausible X-ray IDs were located near bright
stars.  Least-squares fits to sinusoids and the Lomb-Scargle periodogram
were then applied to search for periodic variations in the time series of
the CV candidates. Seven certain periodic variables were confirmed in this
way: W1\opt, W8\opt, W15\opt, W21\opt, W34\opt, AKO~9 and the marginal ID
for W71. We discuss each of these here in turn.

\begin{figure}
\epsscale{0.75}
\plotone{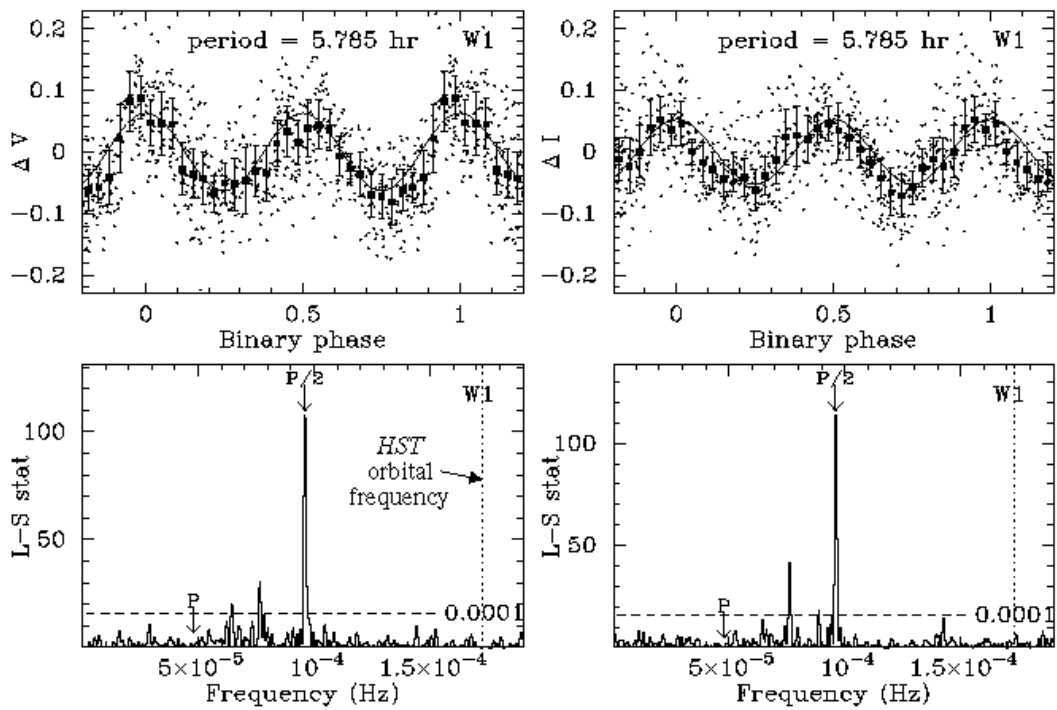}
\caption{Phase plots and Lomb-Scargle power spectra in the $V$ and $I$ band
for W1 (a CV). Individual points and phase bins (with 3-$\sigma$ errors)
are shown for the phase plots and a sinusoid fit is overplotted. The
orbital period (and half of it) are labeled on the power spectra, and the
\hst\ orbital frequency is shown. The horizontal dotted line shows the
power level corresponding to a false-alarm probability of $1\times10^{-4}$.}
\label{fig.w1}
\end{figure}

The ID for W1 is either a 2.89 hr or a 5.78 hr period variable (ellipsoidal
variations). Examination of the light curves folded at the 2.89 and 5.78 hr
periods, and of the power spectra, do not show unambiguous departures from
sinusoidal light curves (Fig.~\ref{fig.w1}), although the $V$-band phase
plot does show some evidence for asymmetry at a little less than the
3-$\sigma$ level, arguing for the longer period. We can use equation
(2.102) from Warner (1995) to place limits on the absolute magnitude of the
secondary:

\begin{equation}
        M_V(2) = 16.7 - 11.1\ {\mathrm l}{\mathrm o}{\mathrm g}\ P_{orb}(h),
\label{eqt.warner}
\end{equation}

where $M_V(2)$ is the absolute magnitude of the secondary and $P_{orb}$ is
the orbital period measured in hours (where 2 hr$\lesssim P_{orb} \lesssim$
10 hr). This relationship is appropriate for field CVs of roughly solar
metallicity. At fixed mass, stars with lower metallicity will have brighter
secondaries (Stehle, Kolb \& Ritter 1997), and so equation~1 gives a lower
limit to the luminosity of secondary stars in 47~Tuc (an upper limit to
\mv).  For W1\opt\ ($M_V=5.86$) using a 5.78 hr period gives a limit of
$M_V=8.2$ from equation~1, while using a 2.89 hr period gives $M_V=11.6$.
The longer period is therefore favored unless the accretion disk completely
dominates in the $V$-band.  This may be possible if W1 is a non-magnetic
nova-like CV, explaining the relatively blue $V-I$ color of W1\opt.
However, the \fxfopt\ value for this system is much more consistent with a
magnetic CV or a DN system (see Fig.~\ref{fig.fxfopt}), and a period of
2.89 hrs would be shorter than any of the UX UMa systems given in
Fig.~\ref{fig.permv}a, and all but one of the DQ Her systems. Modeling of
the accretion disk and secondary in this system may help determine which
period is correct.

Note that both W8\opt\ and W15\opt\ are optically crowded and the absolute
photometry in $V$ and $I$ is relatively poor (an $I$ magnitude could not be
self-consistently derived for W15\opt). However, since this project was
optimised for time series accuracy, the quality of the time series is much
higher than that of the absolute photometry. Both W8\opt\ and W15\opt\ are
clearly eclipsing systems (see Fig.~\ref{fig.w8w15}), with unambiguous
period determinations of 2.86 hr (at the upper edge of the period gap for
field CVs) and 4.23 hr respectively.  Both of the associated X-ray sources
for these CVs are very hard (GHE01a). Detection of these objects as
eclipsing strongly supports the argument of GHE01a that self-absorption by
the accretion disks are responsible for the hard spectra.  See
\S~\ref{sect.permv} for more details about these two binaries.

The ID for W21 is very near to a strong saturation trail and evidence for a
periodic signal was only detected in the single--pixel analysis.
Fig.~\ref{fig.w21} shows the secure 52 minute variation (this may also be a
104 minute orbital period). The lack of strong signal at 104 minutes in the
power spectrum and the large amplitude of the time series (0.28$\pm$0.14)
tentatively favor 52 minutes being the orbital period. Poor phase coverage
at phases 0.4--0.75 makes it difficult to distinguish between these two
periods, although the phase plot does show suggestive evidence of
asymmetry.  In the 52 minute period case the secondary is possibly a
degenerate star since 52 minutes is well below the canonical period minimum
for field CVs of $\sim75$ min, for non-degenerate secondaries. However,
some field CVs with non-degenerate secondaries do have periods about this
small (e.g. a 59 minute orbital period for V485~Cen; Augusteijn et
al. 1996).

The light curve for W34\opt\ is discussed in detail in Edmonds et
al. (2002b).  Although the strong 97.45 minute period is very close to the
\hst\ orbital period (96.5 minutes), Edmonds et al. (2002b) argue that this
is not the result of an artifact showing up as a false signal at the \hst\
orbit.  The binary W36\opt\ (AKO~9) is a long period eclipsing CV with a
subgiant secondary and a period unambiguously determined to be 1.108 days
(see also Edmonds et al. 1996, GHE01a and Knigge et al. 2002). The blue
variable and possible counterpart to W71 has a 1.19 hr power spectrum
signal with a likely 2.37 hr orbital period, based on the small but
significant asymmetry in the phase plot using the longer period (see
Fig.~\ref{fig.w71}). If this is a CV it would fall within the period gap
for field CVs ($2.2\lesssim P_{orb}\lesssim 2.8$hr).  Although this
variable may be a magnetic CV, and the source is relatively soft,
consistent with an AM~Her system, we favor the explanation (supported by
the astrometry), that the \cha\ source is not associated with the nearby
blue, variable star (see \S~\ref{sect.bluevar} for possible explanations of
its nature).

For the other CV candidates, deeper searches for periodic variations were
made and several marginal signals were found (see
Fig.~\ref{fig.marg-tseries}). For V1, the strongest $V$-band power spectrum
peak in a reasonable period range corresponds to an orbital period of
$\sim$7 hr (for ellipsoidal variations). For V2 peaks are found at 13.67/2
hr ($V$) and 6.01/2 hr ($I$) and a strong signal is also seen in $V$ at
1.44 days and a 1.5 day signal in $I$. The shortest of these corresponds to
a plausible orbital period, while the two longest ones are unlikely to be
orbital periods for this apparent MS secondary, unless it is evolved.  For
V3, peaks are found in both $V$ and $I$ corresponding to a possible orbital
period of $\sim$3.7 hr (close to the 3.83 hr period present in the X-ray
light curve; GHE01a). The ID for W45 shows a significant peak at 1.87 hr in
$V$ (possibly a 3.75 hr orbital period). A weaker signal is found in $I$ at
the same period and a strong signal at 1.78 hr.

The ID for W2 shows possible periods at 2.2, 5.9 and 8.2 hr. These differ
from the statistically significant X-ray period of 6.287 hr (see
Fig.~\ref{fig.w2} for an X-ray phase plot and Fig.~\ref{fig.w2-w120v}a for a
$V$-band power spectrum and phase plot). The power spectrum is extremely
noisy, at least partly because the counterpart lies on a diffraction spike
from a nearby giant star. Flickering may be an extra source of noise.  The
ID for W120 shows a marginally significant peak in the $V$ power spectrum
of 1.32 hr, with a false-alarm probability (FAP) of 1.4$\times10^{-2}$ (see
Fig.~\ref{fig.w2-w120v}b). The power spectrum is otherwise very clean, with no
other peaks having a FAP $< 0.9$.  The possible counterpart to W10 has a
2.2 or 4.4 hr period and may also be a CV, but crowding prevents us
obtaining useful limits on the color.

Figure~\ref{fig.rms} shows plots of time series rms vs $V$ and $I$ for the
PC1, WF2 and WF4 chips (the few WF3 variables are shown on WF4). In the
cases where $I$ values are not available (W15\opt, W33\opt, W45\opt\ and
W70\opt) we have assumed that the object is found on the MS in the $V$ vs
$V-I$ CMD.  As expected, significantly higher rms values than average are
found for stars with large amplitude, periodic variations (e.g. W1\opt,
W8\opt\ and W15\opt).  Among the CV candidates, significant non-periodic
variations were found for V1, V2, V3 and W25\opt.  These objects are
clearly visible in the $V$ and $I$ images, and are not affected by
diffraction spikes or saturation trails, so the source of the enhanced
noise for these objects is probably flickering from the hot spot or
accretion disk. Strong temporal changes in the PSFs correlated with
telescope breathing caused likely spurious signals to appear in the power
spectra near the \hst\ orbital frequency (0.1725 mHz) for these four large
amplitude variables.

The same explanation cannot be confidently applied to the possible enhanced
noise for W33\opt, W45\opt, W70\opt\ and W120\opt. These objects are all so
crowded that none of them were detected in the original $V$ and $I$ band
images by AGB01, and the enhanced noise could have some contribution from
this crowding. Strong signals near the \hst\ period are only found for
W33\opt\ in the $I$ band. By contrast, the ID for W44\opt\ is relatively
isolated, though faint, and the higher rms values found here are also
suggestive of non-periodic variations.

\subsubsection{GO-7503 data}

Significant variability was found for several of the CV candidates by
comparing the 3rd epoch images of the \hst\ program GO-7503 with those
obtained in the first two epochs (epoch~1: GO-5912, and epoch~2:
GO-6467). Fig.~\ref{fig.7503-var} shows those cases where F300W variability
was discovered (small amplitude variability was also found for V2, a known
DN, but is not shown here). Most of the variations are obvious from
Fig.~\ref{fig.7503-var} with the exception of W44\opt\ which is 0.36 mag
fainter in the 1st epoch than it is in the 2nd (and 0.30 mag fainter in the
1st than in the 3rd), and W122\opt\ which is 0.87 mag fainter in the 1st
epoch than the 2nd and 0.52 mag brighter in the 2nd epoch than the 3rd.
Some of these variations may represent dwarf nova (DN) outbursts,
especially W51\opt\ and W56\opt. A detailed analysis of the long-term
variability of the CVs will be deferred to a later publication (where these
results will be combined with subsequent \hst\ observations made with ACS).

\subsubsection{Period-\mv\ plot}
\label{sect.permv}

Having a sample of CVs and CV candidates with period and \mv\
determinations at a well-known distance allows us to examine the period
luminosity relationship expected for CVs. Figure \ref{fig.permv}a shows a
plot of period vs \mv\ for CVs and CV candidates in 47~Tuc along with those
of CV1 and CV6 in NGC~6397 (Grindlay et al. 2001b, hereafter GHE01b;
Kaluzny \& Thompson 2002), the DN in M5 (Neill et al. 2002), two CVs in
NGC~6752 (Bailyn et al. 1996) and the magnetic CV in M67 (Gilliland et
al. 1991). The straight line is equation~\ref{eqt.warner} from Warner
(1995), for solar-metallicity stars. This will shift to brighter magnitudes
(by about 1 mag) for metal-poor secondaries, so CVs that are near or above
this line are likely to have evolved secondaries.  Some of the $V$-band
light will be from the accretion disk and hot spot (as well as the
secondary) and this will also push the $V$ magnitudes to brighter values
(to the left in \ref{fig.permv}a). Figure~\ref{fig.permv}b shows the plot
for the blue variables described earlier.

\begin{figure}
\epsscale{0.75}
\plotone{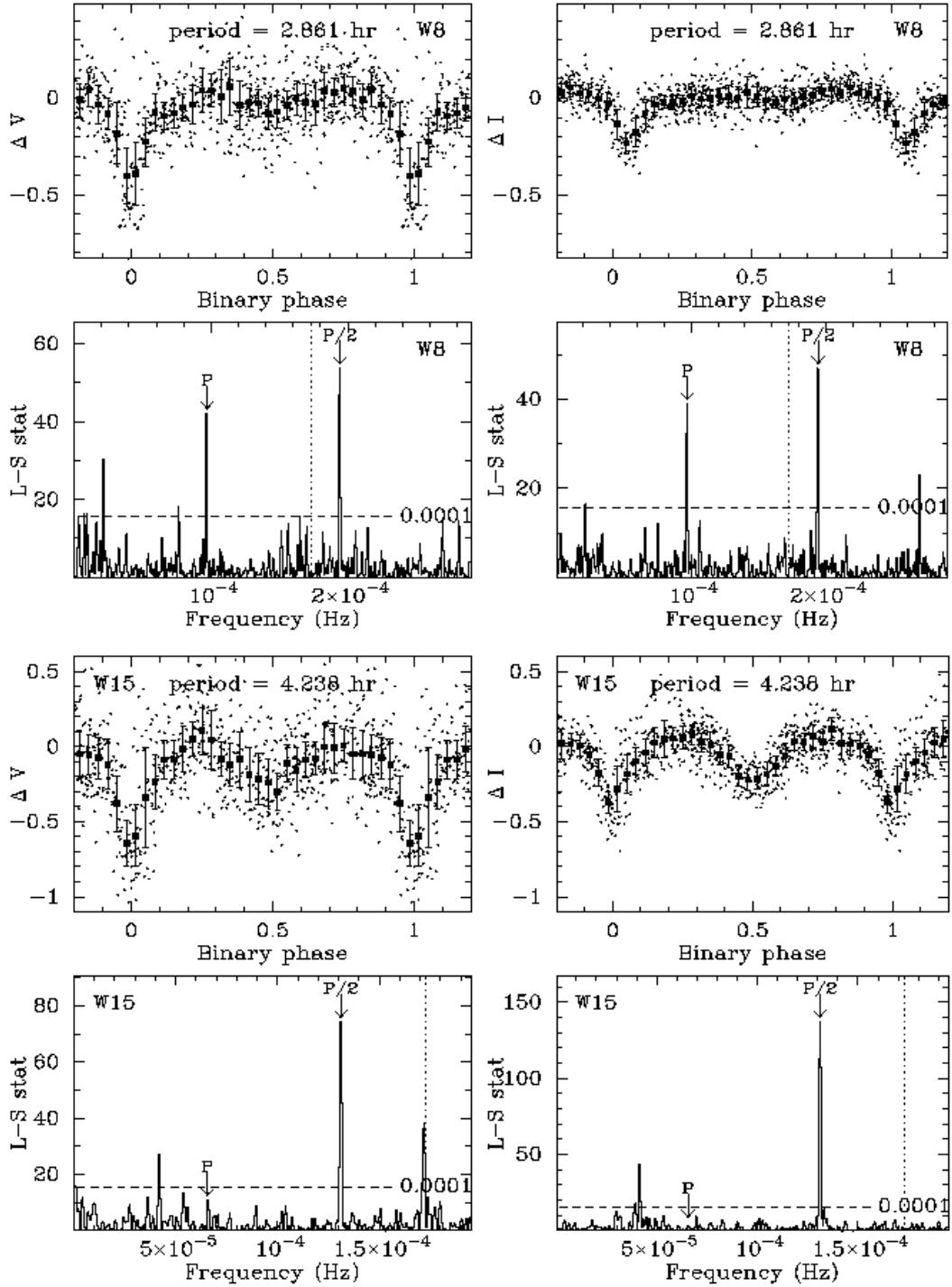}
\caption{Phase plots and power spectra in the $V$ and $I$ band for the
eclipsing CVs W8\opt\ and W15\opt. Note, for both systems, the deeper
eclipses in the $V$ band compared to the $I$ band because the blue
accretion disk is being eclipsed. These CVs both have very hard X-ray
spectra, likely caused by significant X-ray absorption by the accretion
disk.}
\label{fig.w8w15}
\end{figure}

\begin{figure}
\epsscale{0.6}
\plotone{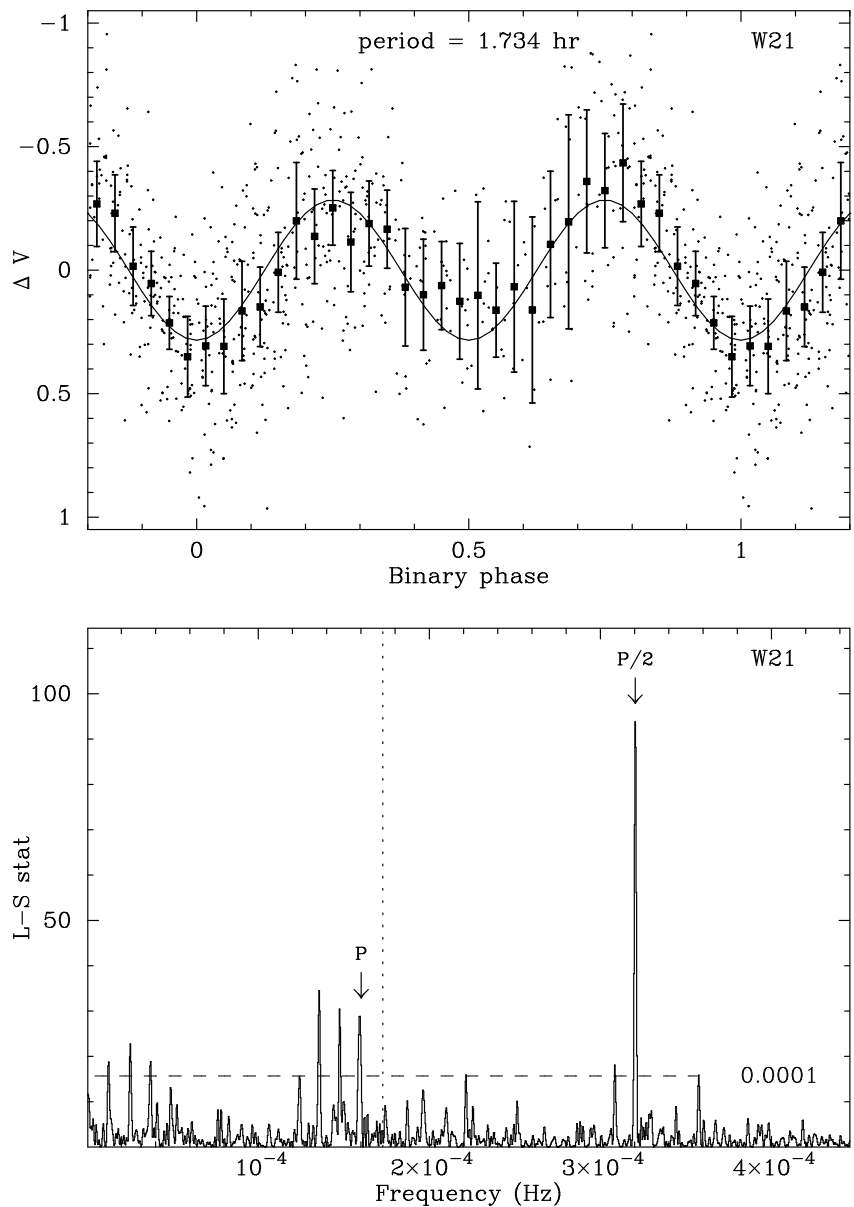}
\caption{The phase plot and power spectrum in the $V$ band for W21\opt,
based on single--pixel photometry (because of saturation no useful data is
available for $I$). We have assumed that the orbital period is 1.734 hr
(for ellipsoidal variations), but the orbital period could also be 1.734/2
hr.}
\label{fig.w21}
\end{figure}

\begin{figure}
\epsscale{0.75}
\plotone{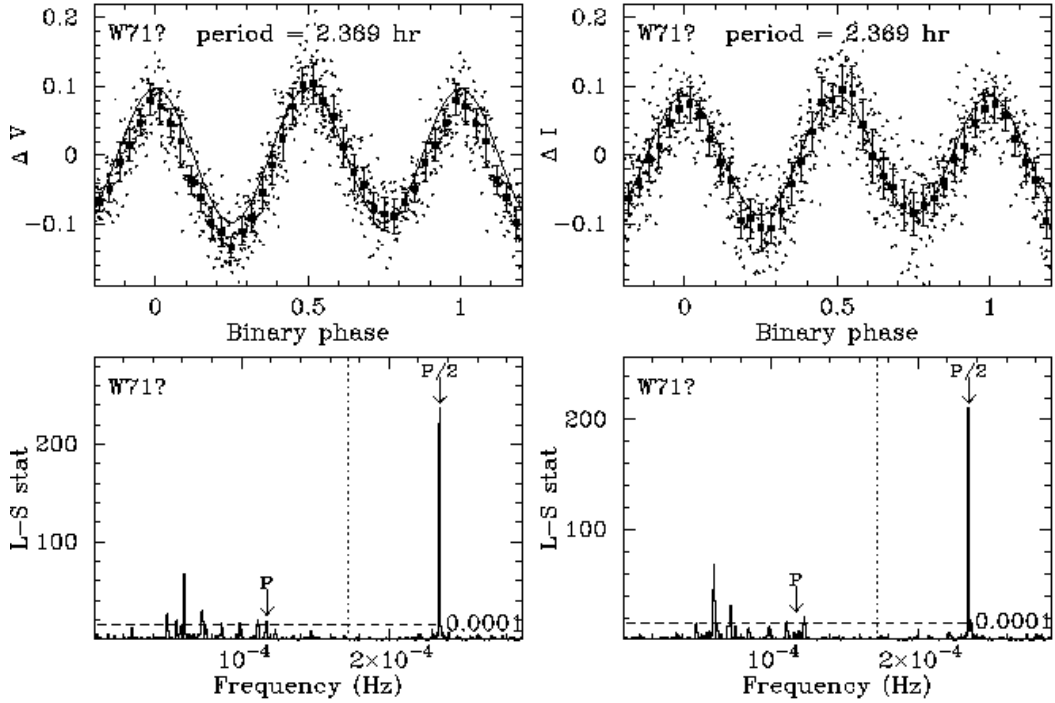}
\caption{The phase plot and power spectrum for a blue variable and marginal
optical counterpart to W71. There are small, but significant deviations of
the phase plot from a sinusoidal curve (overlaid).}
\label{fig.w71}
\end{figure}

\begin{figure}
\plotone{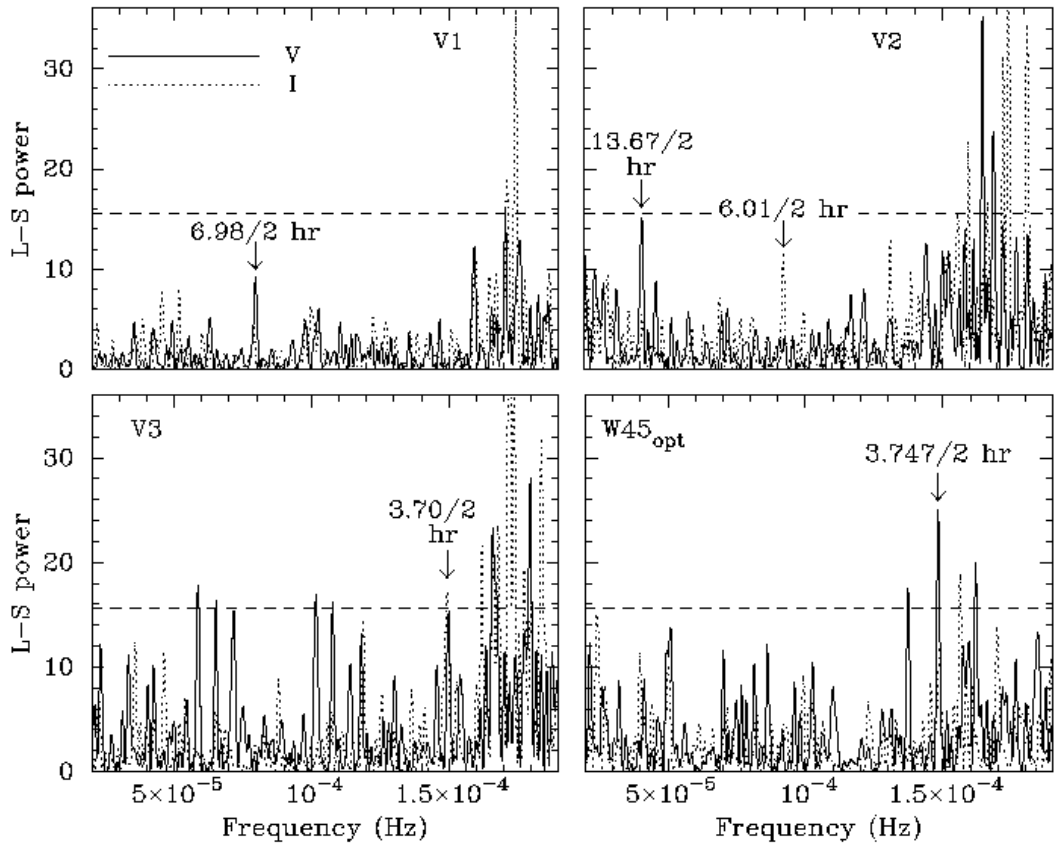}
\caption{Power spectra in the $V$ and $I$ band for V1, V2, V3 and
W45\opt. Possible periods are labeled (for ellipsoidal variations the
periods will be twice the values shown). The X-ray counterpart to V3 (W27)
has a period of 3.83 hr (GHE01a).}
\label{fig.marg-tseries}
\end{figure}

\begin{figure}
\plotone{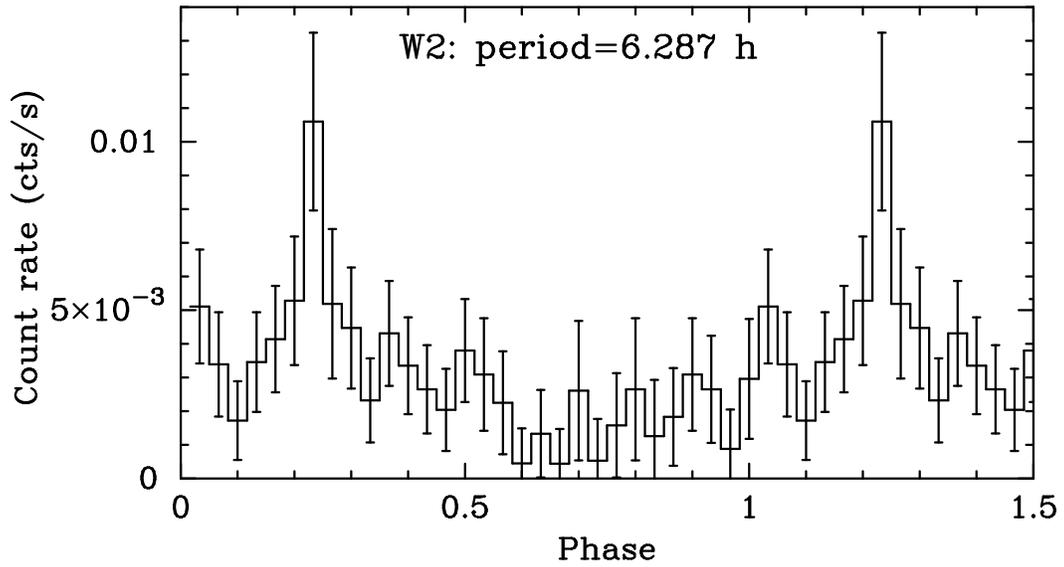}
\caption{\cha\ phase plot for W2 (a CV), folded at the period measured from
the power spectrum, with 1-$\sigma$ error bars.}
\label{fig.w2}
\end{figure}

\begin{figure}
\epsscale{0.75}
\plotone{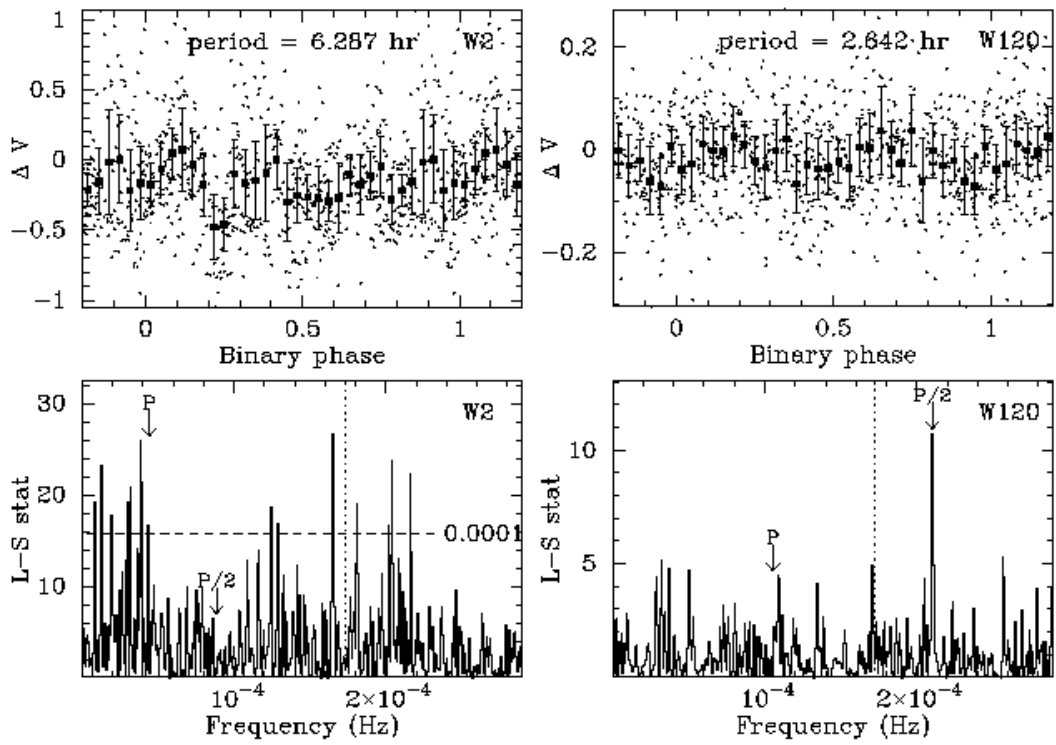}
\caption{The phase plots and power spectra for the CVs W2\opt\ and
W120\opt. For W2\opt\ the phase plot is folded at the X-ray period, and
this period is labeled in the power spectrum.}
\label{fig.w2-w120v}
\end{figure}

\begin{figure}
\epsscale{1.0}
\plotone{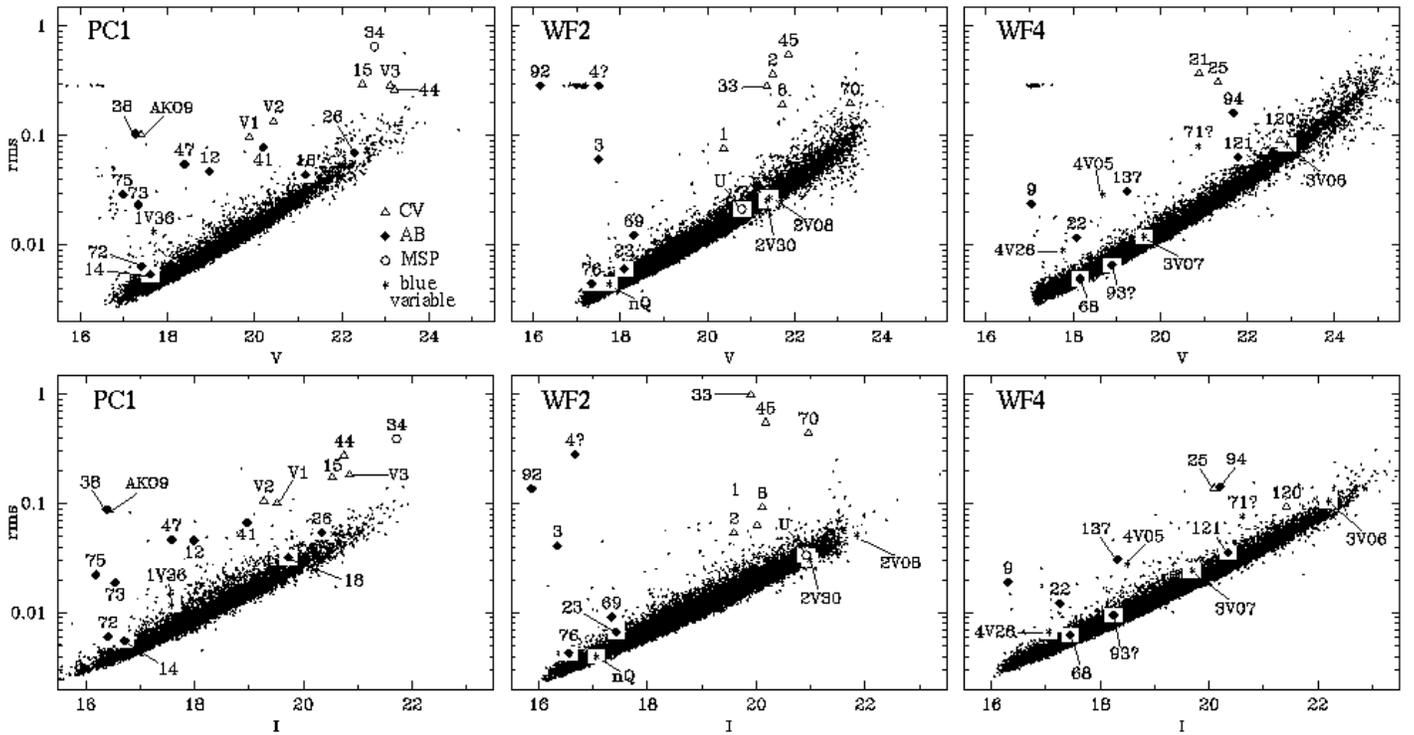}
\caption{Plots of time series rms vs $V$ and $I$ for optical counterparts
to X-ray sources in the GO-8267 data, plus the blue variables discussed in
\S~\ref{sect.bluevar}. The WF3 data (containing relatively few variables)
is not shown, but 3V06 and 3V07 are included with the WF4 CMD. The clump of
bright objects with rms values of $\sim$0.3 are saturated stars in the
$V$-band. The labeling shown in the PC1 plot for $V$ is the same as that
used in the CMDs of Paper~I (triangles = CVs; circles = MSPs; diamonds =
active binaries; `X' = qLMXB; asterisk = unknown). Subsequent figures use
the same labeling.}
\label{fig.rms}
\end{figure}

\begin{figure}
\epsscale{0.8}
\plotone{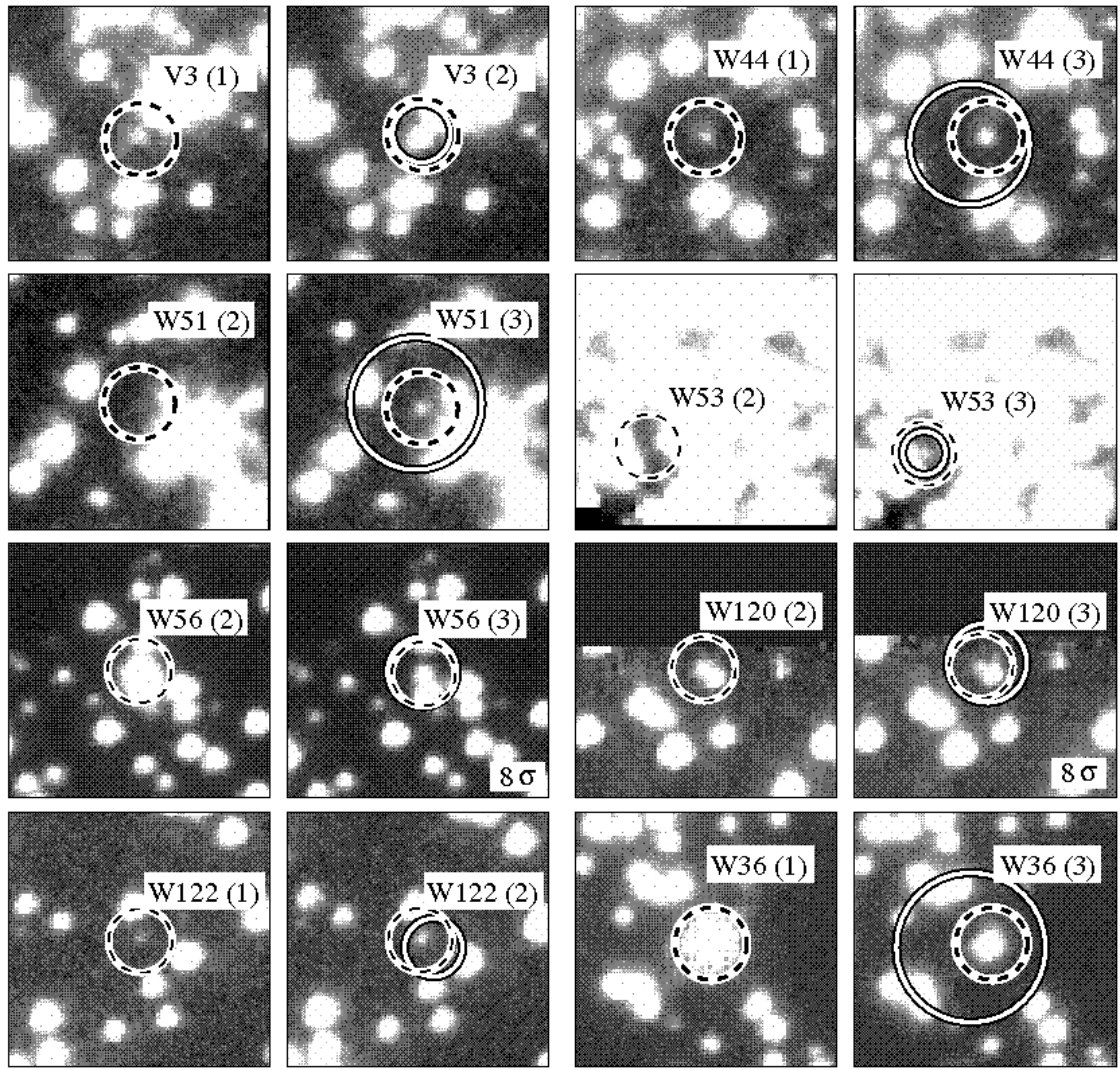}
\caption{Finding charts (in F300W) for the CVs or CV candidates found in
the GO-7503 FoV that show detectable or significant long-term
variability. Data from GO-5912 (epoch~1), GO-6467 (epoch~2) and GO-7503
(epoch~3) are shown (the appropriate epoch number is shown in
parentheses). The solid lines show the 4-$\sigma$ error circles (except in
the cases of W56 and W120 where 8-$\sigma$ error circles are shown for
clarity) and the dashed circles encircle the IDs (with radius 0\farcs2 for
PC1 and 0\farcs1 for the WF chips).}
\label{fig.7503-var}
\end{figure}

\begin{figure}
\epsscale{0.7}
\plotone{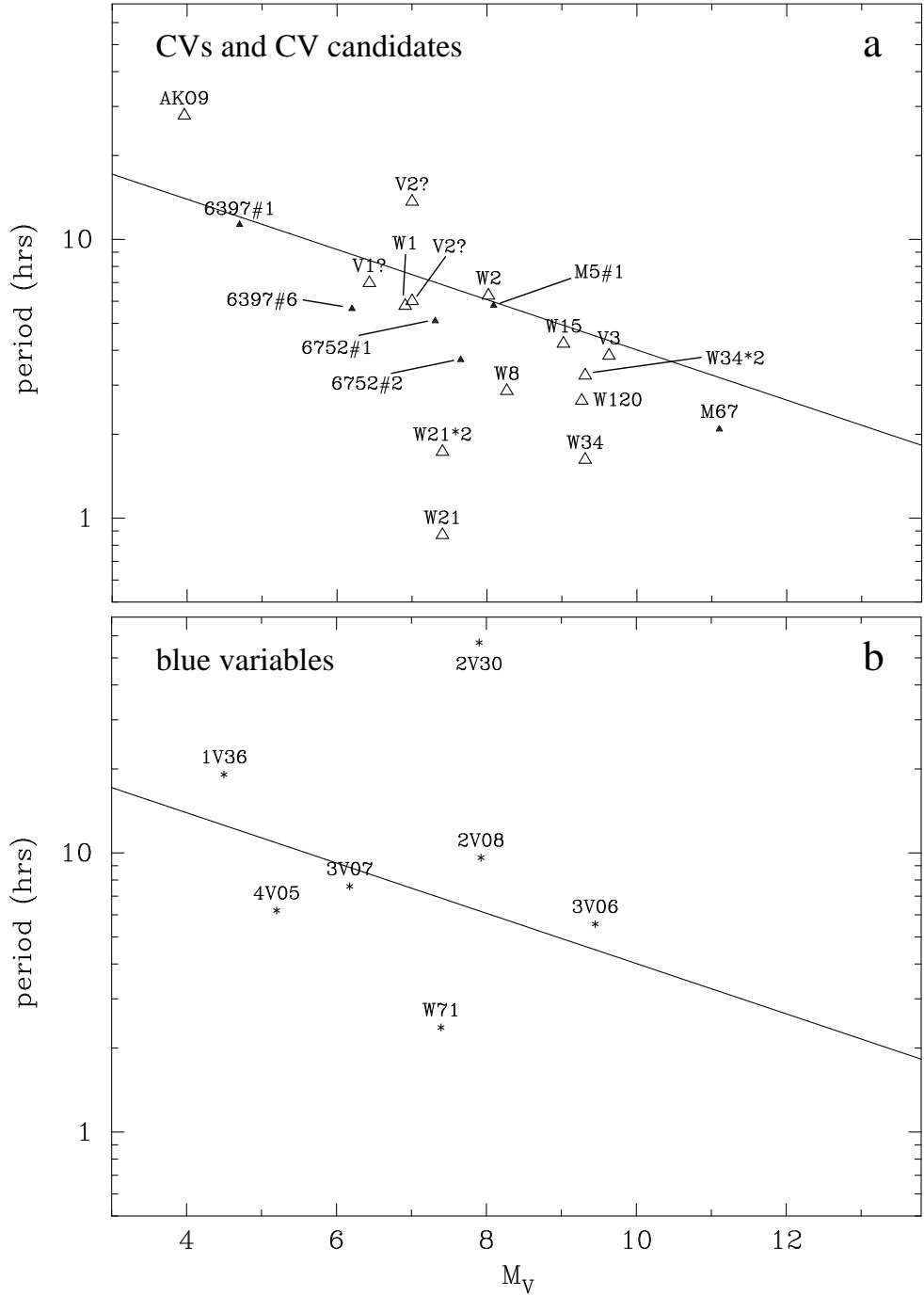}
\caption{A plot of orbital period vs \mv\ for 47~Tuc CVs and CV candidates
(Fig.~\ref{fig.permv}a) and blue variables (Fig.~\ref{fig.permv}b). In
Fig.~\ref{fig.permv}a two CVs from each of the globular clusters NGC~6397
(CV1 and CV6; Grindlay et al. 2001b and Kaluzny \& Thompson 2002) and
NGC~6752 (Bailyn et al. 1996) are shown plus a CV from M5 (Neill et
al. 2002) and M67 (Gilliland et al. 1991). Small filled triangles are used
to plot the non 47~Tuc CVs.  The periods for the blue variables have been
assumed to be twice those found from the peak in the power spectrum. The
straight line shows equation~\ref{eqt.warner} from Warner (1995).}
\label{fig.permv}
\end{figure}

The likely CV W2\opt\ and the CV candidate V3 both lie close to or above
the MS ridge-line in the $V$ vs $V-I$ CMD (see the CMD for PC1 in Paper~I),
showing that the secondary dominates the optical light. As expected these
objects lie close to the Warner relationship. The secondary for W2\opt\
might be evolved, but better photometry is needed to make this conclusion
secure.  The CVs V1 and W1\opt\ fall below equation~\ref{eqt.warner}
because the accretion disk and hot spot (and possibly the WD) make
significant contributions to the light (the CMDs in Paper~I show that the
$V-I$ colors of W1\opt\ and V1 are well to the blue of the MS). The light
curve of the eclipsing CV W15\opt\ (Fig.~\ref{fig.w8w15}) implies that the
accretion disk and hot spot are responsible for $\sim60$\% of the $V$-band
light, and increasing $V$ by $\sim$1 mag to isolate the MS component brings
W15\opt\ into excellent agreement with equation~\ref{eqt.warner}. For
W8\opt\ around 40\% of the $V$-band light appears to come from a source
other than the secondary.  Given the similarity of the light contributions
from the secondaries, equation~\ref{eqt.warner} implies that the W8\opt\
secondary should be about 2 mags fainter than the W15\opt\ secondary, but
instead it is almost 1 mag {\it brighter}.  Possible explanations are that:
(1) we are seeing a grazing eclipse with the inclination slightly less than
90 degrees, (2) a faint MS star is contaminating the light from the CV, or
(3) the normalization for W8\opt\ is incorrect because of the crowding.
Examination of Fig.~\ref{fig.w8w15} shows that the ellipsoidal variations
in W15\opt\ are close to the maximum value expected (see Russell 1945), but
are around a factor of 2 smaller in W8\opt, consistent with any of these
three explanations.  Similar (but less extreme) issues exist for the two
NGC~6752 CVs (both of these systems lie close to the MS in the $V$ vs $V-I$
CMD). The good agreement of the M5 DN with equation~\ref{eqt.warner}
implies that the $V$-band light must be dominated by the secondary.

The variable W34\opt\ has colors that are consistent with those of a CV
($U-V= -0.12$; $V-I$=1.03; \mv=9.3) and appropriately falls below
equation~\ref{eqt.warner}. However, as noted by Edmonds et al. (2002b), the
relatively large amplitude variation of this object in both $V$ and $I$
makes an MSP another possibility, since the light curve is similar to that
of W29\opt (47~Tuc~W). Either possibility is consistent with the X-ray
properties of this object. The variable W21\opt\ ($U-V$=0.15; $V-I$=0.27;
\mv=7.44) is also difficult to categorize. Assuming that the orbital period
is twice the observed period, the star still falls well below
equation~\ref{eqt.warner}. Could an accretion disk for a CV with a period
of 1.73 hr and an expected \mv(2)=14 be this bright? Figure 9.8 of Warner
(1995) shows that the 13 DNe with periods $\lesssim$2 hr mostly have \mv
(disk)= 8--9, with 3 having \mv (disk) $\lesssim$ 7.1. However, we would
expect a very blue $U-V$ color if the disk is this bright.  Alternatively,
the binary could be a WD-WD CV or a WD-NS system in quiescence. Finally,
AKO~9 with its subgiant secondary is well outside the range of
applicability of equation~\ref{eqt.warner}.

\begin{figure}
\plotone{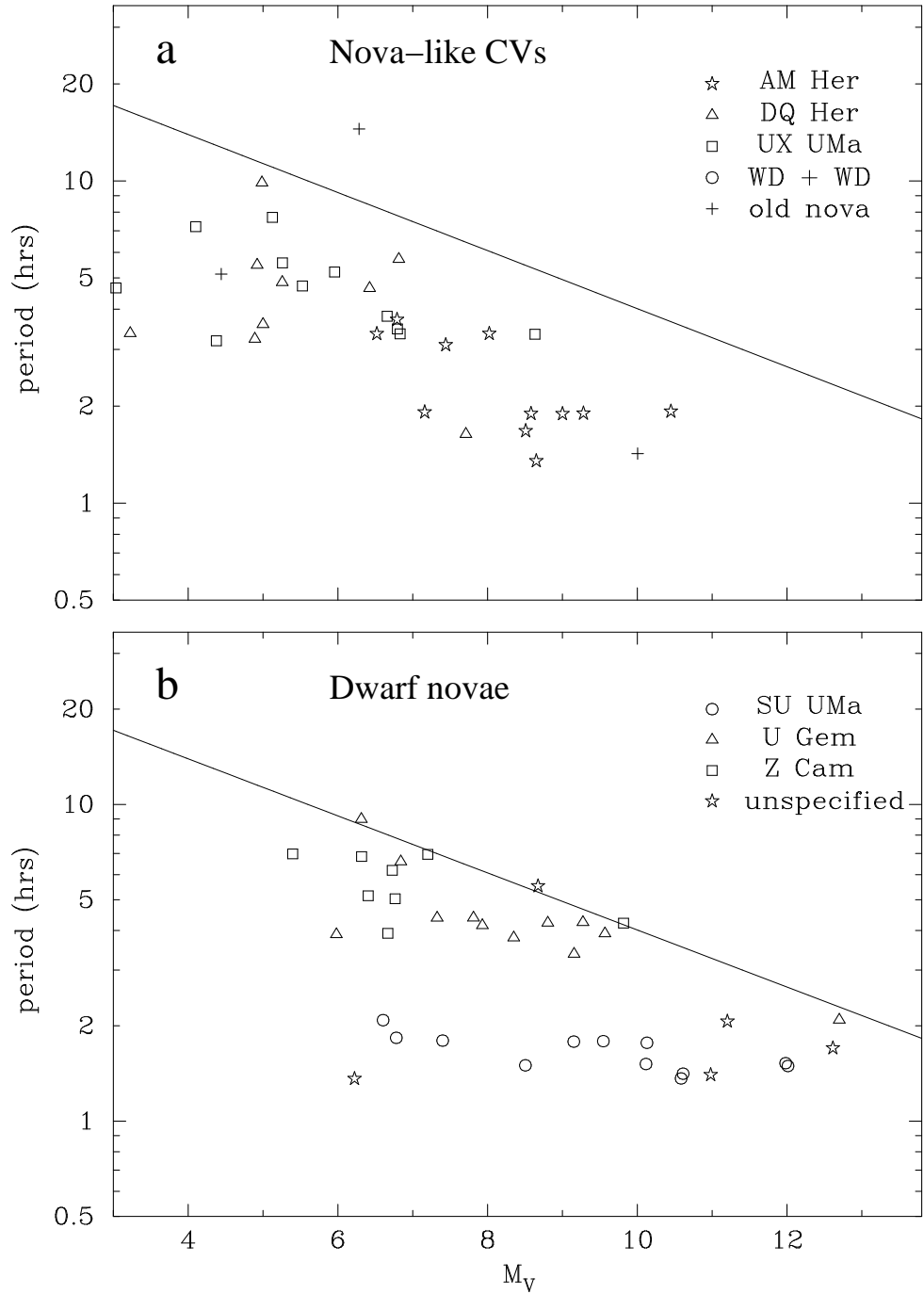}
\caption{A plot of orbital period vs \mv\ for field CVs from
VBR97. Fig.~\ref{fig.permv-ver}a shows nova-like systems and
Fig.~\ref{fig.permv-ver}b shows DN systems.}
\label{fig.permv-ver}
\end{figure}

Figure \ref{fig.permv-ver} shows a period vs \mv\ plot (similar to
Fig.~\ref{fig.permv}) for the field CVs with estimated distances in the
ROSAT study by VBR97. The orbital periods have been taken from Ritter \&
Kolb (1998). The nova-like CVs (systems that are magnetic or do not show
outbursts) and the DNe (outbursting systems) have been separated into two
plots to show systematic differences between these two general classes of
CV.  The labeling of VBR97 for the different CV types is used. The
nova-like CVs are generally further away from equation~\ref{eqt.warner}
than the DNe, showing that the relative light contributions from the
accretion disk and hot spot (compared to the secondary) are larger than for
the DNe. The 47~Tuc CVs (and the other cluster CVs) appear to be more
consistent with the DNe from VBR97 than the nova-like systems.

\begin{figure}
\epsscale{0.75}
\plotone{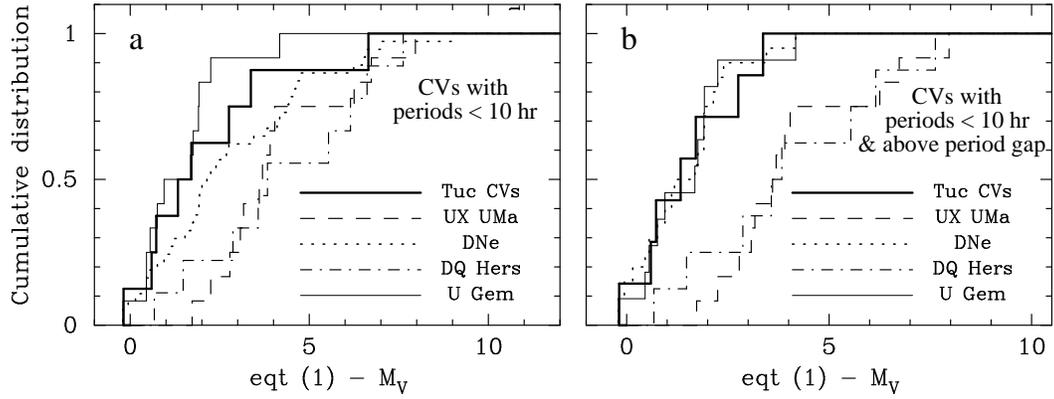}
\caption{Cumulative plots of the difference between \mv\ predicted by
equation~\ref{eqt.warner} (using the binary period) and the measured \mv\
for 47~Tuc CVs (labeled as `Tuc CV's) compared to field CVs from
VBR97. This difference is an estimate of the brightness of the accretion
disk and hot spot, plus (for faint systems) the WD, where higher accretion
luminosities are towards the right of the figure. The secondaries in the
47~Tuc systems will be brighter than solar metallicity secondaries, and so
the 47~Tuc accretion luminosities will be even fainter than shown by these
figures.  The `DNe' distribution is for all DN systems.
Fig.~\ref{fig.cumul-permv}a shows all systems with periods less than 10 hr
and Fig.~\ref{fig.cumul-permv}b shows systems with periods less than 10 hr
but above the CV period gap (this shows systems where the contribution from
the white dwarf is likely to be smaller).}
\label{fig.cumul-permv}
\end{figure}

By calculating the magnitude predicted by equation~\ref{eqt.warner} (using
the binary period) and subtracting \mv, we quantify the offsets from
equation~\ref{eqt.warner} and examine the light contribution from sources
other than the secondary (mainly the accretion disk, hot spot and WD). This
procedure will overestimate the contribution from the disk and other
non-secondary components, since equation~\ref{eqt.warner} underestimates
the contribution from a metal-poor secondary.  Figure \ref{fig.cumul-permv}
shows the cumulative distributions of the offsets from
equation~\ref{eqt.warner} for the 47~Tuc and the field CVs. We have
eliminated binaries with periods $>$10 hr to remain consistent with the
period limitations on equation~\ref{eqt.warner}. Figure
\ref{fig.cumul-permv}a shows all such systems and
Fig.~\ref{fig.cumul-permv}b shows only systems with periods above the CV
period gap (the latter figure avoids the shortest period systems likely to
have a greater light contamination from the WD).  For systems above the
period gap, the offset distributions for the 47~Tuc CVs and the nova-like
systems are distinguishable at the 99.55\% level using the KS-test. A
subset of the nova-like systems, the DQ~Hers, are distinguishable from the
47~Tuc CVs at the 93.8\% level (the nova-like and DQ~Her offsets are
distinguishable from each other at the 1.2\% level).  By contrast, the
47~Tuc CV offsets are distinguishable from the DNe offsets and the U~Gem
systems at only the 1.4\% and 0.9\% levels respectively.  This suggests
that the 47~Tuc CVs have accretion luminosities that are even lower than
those found in U~Gem systems and DNe since, as mentioned above, the
non-secondary component in the 47~Tuc CVs is being underestimated
here. This then suggests that the 47~Tuc CVs have lower accretion rates.
More evidence supporting this conclusion is given in \S~\ref{sect.fxfopt},
and discussion is given in \S~\ref{sect.cvdisc}.

\subsection{Blue Variables}
\label{sect.bluevartseries}

\begin{figure}
\epsscale{1.0}
\plotone{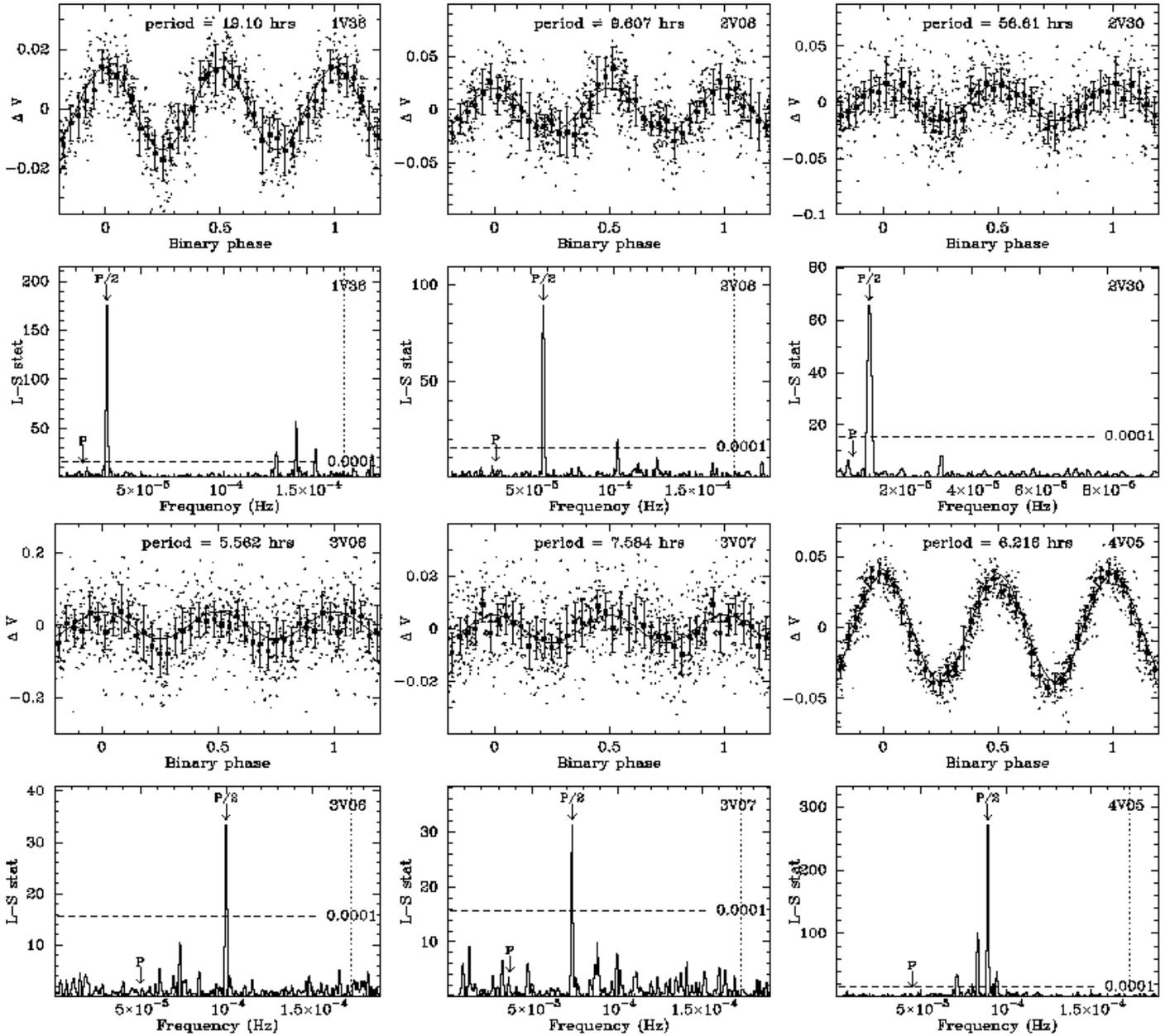}
\caption{Phase plots and power spectra in the $V$-band, for the blue
variables from AGB01. In each case the orbital period has been assumed to
be twice that derived from the peak in the power spectrum. Sinusoidal fits
to the data are shown. The marginal optical counterpart to W71, another
blue variable, was previously shown in Fig.~\ref{fig.w71}.}
\label{fig.bluevar-v}
\end{figure}

As described in Paper~I, several blue, variable stars discovered by AGB01
are not obviously associated with X-ray sources and have colors that are
different from most cluster and field CVs.  The time series for these blue
variables are shown in Fig.~\ref{fig.bluevar-v} and the period versus \mv\
plot is shown in Fig.~\ref{fig.permv}b. The variable 1V36 has a sinusoidal,
9.55 hr variation; 2V08 and 2V30 have 4.8 hr and 28.3 hr variations, and
3V06, 3V07 and 4V05 have 2.78 hr, 3.79 hr and 3.11 hr variations
respectively. Note that the absolute magnitudes of the secondaries (if MS
members) are likely to be significantly fainter than the observed \mv\
because these stars are all blue in $V$ vs $V-I$.  We have assumed that the
orbital period is twice the detected period, as appropriate for most CVs
(Roche-lobe filling, MS secondaries with ellipsoidal variations) with
periodic variations detected in the optical. If this assumption is correct,
then 1V36, 2V08, 2V30 and 3V06 are unlikely to be filling their Roche lobes
since they fall above equation~\ref{eqt.warner} in Fig.~\ref{fig.permv}b,
and are therefore unlikely to be CVs (as already noted for 1V36 by AGB01).
If any of the secondaries are evolved then they could fill their Roche lobe
at the observed period, but such secondaries should be unusually red, and
therefore to explain the observed blue $V-I$ colors we probably require an
unusually UV-bright, luminous component. Even if evolved companions were
expected to give unusually bright disks, only 2V08 and 2V30 have $U-V$
colors as blue as most of the other CVs, so the blue variables do not seem
to have unusually bright disks.  The blue color for 3V07 implies that only
a small fraction of the optical light comes from a MS star, and therefore
the secondary for this system is also likely to fall above the Warner
relationship (the \mv\ offset of 3V07 from equation~\ref{eqt.warner} in
Fig. \ref{fig.permv} of $\sim$1 mag roughly equals its offset from the MS
in the $V$ vs $V-I$ CMD.

Another possibility is that the orbital periods are the same as the
observed periods, and that the variations are caused by heating of one side
of an MS star. In this case some of the blue variables (like 1V36) may not
have to contain evolved secondaries for Roche lobe filling to be taking
place. There are at least two problems with this explanation. Firstly, a
UV-bright, relatively luminous component is probably required to cause
measurable heating effects, but only 2V08 and 2V30 have $U-V$ colors that
are comparable to those of the bluest CVs.  Secondly, because sinusoidal
signals with low amplitudes (either from heating or ellipsoidal variations)
are generally not detectable even in the brightest CVs because of noise
from flickering. If the blue variables are CVs then we would expect
flickering to mask the small sinusoidal signals that are observed. Instead,
the rms values for the blue variables are quite low (Fig.~\ref{fig.rms})
implying low amplitude flickering, despite having bluer colors and
(presumably) relatively bright disks, if they are CVs.

We note that these results are unlikely to be some sort of photometric
artifact (such as the effects of bad pixels), since several of these
objects exist and none of them show unusual PSF fits, or excessive levels
of crowding. Also, the power spectra show little if any extra power near
the \hst\ orbital frequency. See Sect. \ref{sect.bluevar} for more
discussion about these unusual objects.

\subsection{Active binaries}
\label{sect.abtseries}

\begin{figure}
\epsscale{0.7}
\plotone{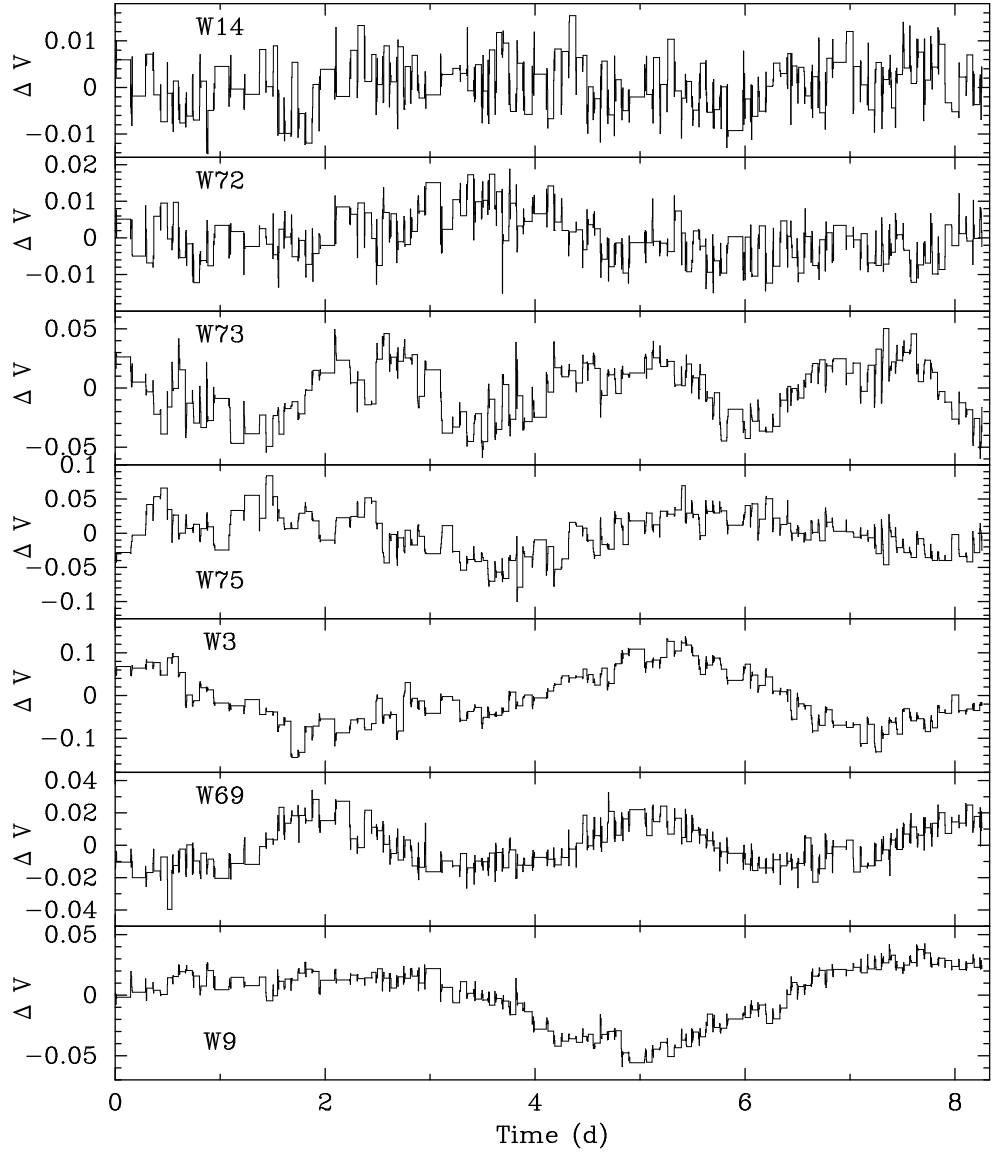}
\caption{Time series for a sample of active binaries with long
periods. These are a mixture of red stragglers (W3\opt\ and W72\opt) and
red straggler candidates (W14\opt), binaries and likely BY~Dra variables
located near or above the MSTO (W9\opt, W73\opt\ and W75\opt) and a
variable located well above the MS ridge-line (W69\opt). Short horizontal
segments in the time series show various gaps in the time series.}
\label{fig.bydra}
\end{figure}

\begin{figure}
\epsscale{0.75}
\plotone{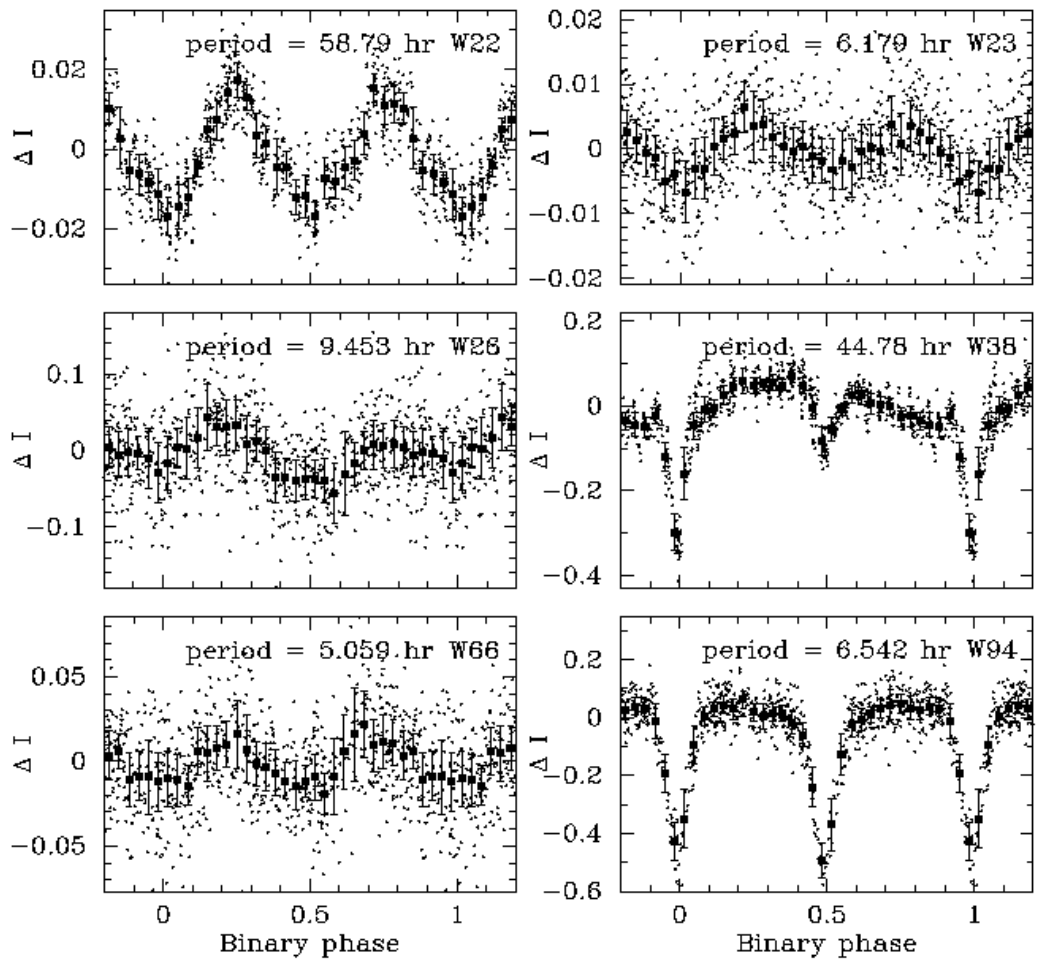}
\caption{Phase plots for a sample of active binaries not found by AGB01
(because of crowding), plus the faint variable W66\opt.}
\label{fig.other-abs}
\end{figure}

The phase plots for W12\opt, W92\opt, W137\opt\ and W182\opt\ (eclipsing
binaries) and W41\opt, W47\opt\ and W163\opt\ (contact binaries) are shown
in AGB01. Light curves for most of the BY~Dras and red stragglers from
AGB01 that have also been identified as active binary candidates are shown
in Fig.~\ref{fig.bydra}, and phase plots for a selection of active binary
candidates not found by AGB01 are shown in Fig.~\ref{fig.other-abs}.  Note
that ambiguity exists in the period determination for several of the
variables shown in Fig.~\ref{fig.other-abs} (especially for W22\opt\ and
W94\opt). In these cases we have assumed that the orbital period is twice
the observed period. Clearly, W38\opt\ and W94\opt\ are eclipsing binaries,
and the other new objects are a likely mixture of BY~Dra and W~UMa systems
(W66\opt\ was classified by AGB01 as a non-eclipsing contact binary).

\section{Analysis}
\label{sect.anal}

\subsection{Radial Distribution of Sources: Comparison Between Source Classes}
\label{sect.rad}

\begin{figure}
\epsscale{0.75}
\plotone{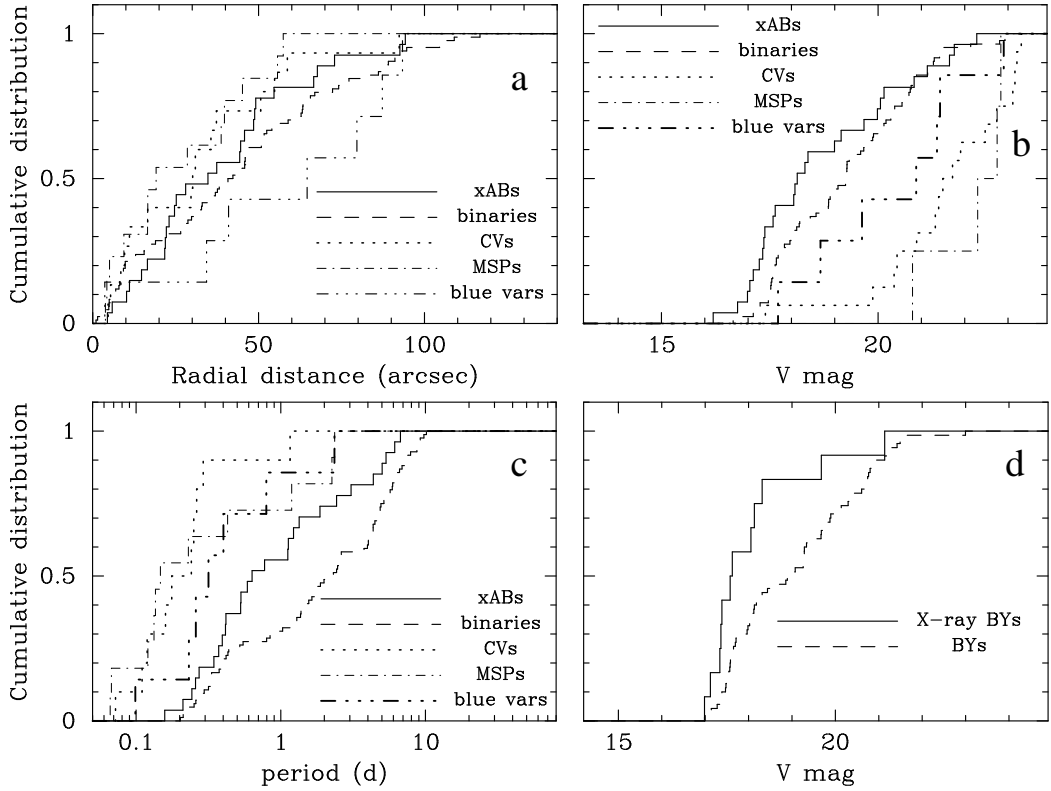}
\caption{Cumulative radial distributions for several classes of 47~Tuc
binary: CVs, MSPs, binaries detected by AGB01 (`binaries'), the subset of
these binaries (`xABs') detected in X-rays and the blue variables
(Fig.~\ref{fig.radall}a); The $V$ distribution of this same subset of
binaries, where the MSP list includes only 47~Tuc~U, 47~Tuc~W (W29), W34
and the possible ID for 47~Tuc T (Fig.~\ref{fig.radall}b); The period
distribution for these binaries, where the periods are derived from AGB01,
Camilo et al. (2000) and this work (Fig.~\ref{fig.radall}c); The $V$
distribution for the AGB01 binaries classified as BY~Dra systems and the
subset of these systems that are X-ray sources (Fig.~ \ref{fig.radall}d).}
\label{fig.radall}
\end{figure}

Here, we compare the radial offsets, with respect to the center of 47~Tuc
(`radial distributions') of the different classes of binary.  Although the
GO-7503 dataset has uncovered a number of likely CVs and at least 2 active
binary candidates, for radial distribution analysis we restrict ourselves
to the GO-8267 dataset because it contains a much more complete sample of
active binary candidates. Figure \ref{fig.radall} shows the radial
distributions of: (1) the radio detected MSPs (plus W34), (2) the CVs, (3)
the binaries discovered by AGB01 (with the relatively small fraction of
X-ray bright systems and the two confirmed CVs removed), (4) the candidate
active binaries or `xABs' (MS or red straggler binaries either from AGB01
or this work with apparent X-ray counterparts) and (5) the blue
variables. Note the similarity between the CV and the MSP radial
distributions (KS-test probability = 14.0\%). This similarity is not
surprising because the single MSPs should have masses of $\sim1.4$\mdot\
and the MSP binaries only $\sim$0.02--0.2\mdot\ more. The detected CV
candidates should have masses ranging from $\sim$0.95\mdot\ for faint
objects like V3 and W120\opt\ ($\sim$0.4\mdot\ secondary and
$\sim$0.55\mdot\ primary) up to $\sim$1.6 \mdot\ for AKO~9 ($\sim$0.8\mdot
primary and secondary). The 7 blue variables are slightly less centrally
concentrated than the CVs (KS-test probability = 88.2\%) and the MSPs
(KS-test probability = 93.8\%). Eliminating 2V08 and 2V30, the two
variables with colors most like white dwarfs or CVs, causes the above
KS-test probabilities to drop to 53\% and 55\%, suggesting that the blue
variables are heavy objects (2V08 and 2V30 are relatively
distant). However, the remaining sample of 5 is very small for statistical
tests.

The radial distribution of the AGB01 binaries is slightly less centrally
concentrated than those of the CVs and MSPs (KS-test probabilities of 77\%
and 81\% respectively). These results are only marginally statistically
significant, but are a likely result of lower masses for some of the AGB01
binaries, especially the faintest ones. The active binaries appear to be
marginally more centrally concentrated than the AGB01 binaries, but the
KS-test probability (45\%) is not statistically significant.  If real, this
difference likely results from the generally brighter $V$ magnitudes (and
higher masses) of the active binaries compared to the AGB01 binaries (see
Fig \ref{fig.radall}b), since the $V$ distributions differ at the 90\%
probability level. Also, the $V$ magnitudes of the X-ray bright BY~Dras are
brighter than the $V$ magnitudes of the much larger BY~Dra population from
AGB01 (see Fig \ref{fig.radall}d; KS test probability = 93.5\%). The active
binary periods extend to higher values than the periods of the CVs and the
MSPs (Fig.~\ref{fig.radall}c), and most strikingly they are significantly
shorter than those of the AGB01 binaries (KS probability = 98.9\%).

\begin{figure}
\plotone{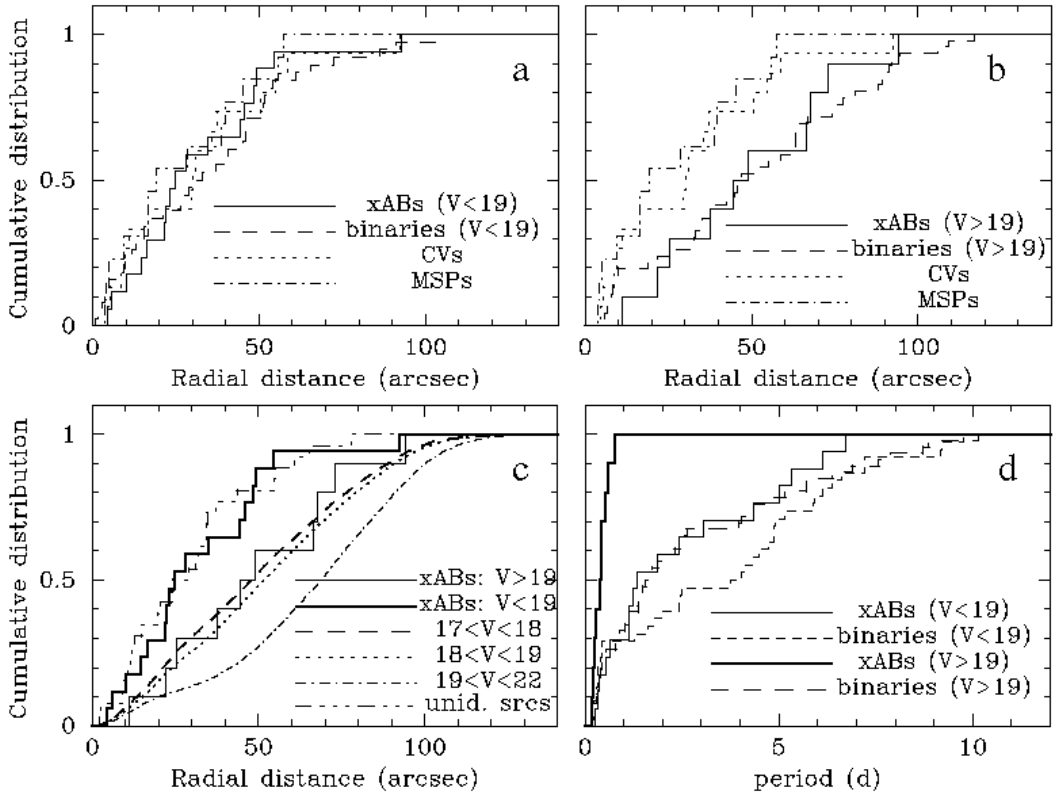}
\caption{Radial distributions for bright ($V<19$; Fig
\ref{fig.rad-bright-faint}a) and faint ($V>19$; Fig
\ref{fig.rad-bright-faint}b) subsets of the AGB01 binaries and the active
binaries. Fig.~\ref{fig.rad-bright-faint}c shows the radial distributions
of the bright and faint active binaries, the total stellar population
(mostly single stars) in the quoted $V$ ranges, and the unidentified X-ray
sources in the GO-8267 field of view; Fig.~\ref{fig.rad-bright-faint}d
shows the period distributions for the bright and faint AGB01 binaries and
the active binaries.}
\label{fig.rad-bright-faint}
\end{figure}

We now consider the AGB01 binaries and active binaries with $V<19$ and
$V>19$ (this divides the AGB01 binaries into two evenly sized groups).  In
the bright subsample the radial distributions of the AGB01 binaries and the
active binaries are indistinguishable from each other (KS-test probability
= 16\%).  The radial distributions of both the AGB01 and the bright active
binaries are indistinguishable from those of the MSPs (KS-test probabilities
= 30\% and 29\%) and CVs (KS-test probabilities = 16\% and 9\%; see
Fig.~\ref{fig.rad-bright-faint}a).

Considering the fainter binaries ($V>19$), the radial distribution of the
faint AGB01 binaries and the faint active binaries are indistinguishable
from each other (KS-test probability = 12\%). The radial distributions of
both the faint AGB01 binaries and the faint active binaries (see Figure
\ref{fig.rad-bright-faint}b) are less centrally concentrated than those of
the MSPs (KS-test probabilities = 97.0\% and 83\%), with similar results
for the CVs (KS-test probabilities = 93\% and 85\%). These results are
consistent with the faint active binaries being less massive than the
bright active binaries.

Figure \ref{fig.rad-bright-faint}c shows the radial distribution of the
sources in the GO-8267 FoV that have not been identified, plotted with the
bright and faint active binaries and the general stellar population in
three different $V$-band ranges. The unidentified sources are
indistinguishable from the bright active binaries, the CVs and the MSPs,
showing that they are consistent with any of these groups, or a combination
of all three types. The faint active binaries have a radial distribution
that is almost identical to that of the stars with $17<V<18$ (probability =
0.73\%), and is considerably more concentrated than the stars with
$19<V<22$ (probability = 84.7\%). All of these results are consistent with
expectations for mass segregation of binaries and stars in 47~Tuc, and with
the statement of AGB01 that incompleteness as a function of radius and
magnitude is expected to be small even for stars as faint as V$\sim$21.5.

The most striking result is that the period distribution of the faint
active binaries differs from that of the faint AGB01 binaries at the
99.99\% probability level. Figure \ref{fig.rad-bright-faint}d shows that
while {\it all} of the faint active binaries have periods $\lesssim$0.7
days, only $\sim$30\% of the AGB01 binaries have periods below this
value. We discuss in \S~\ref{sect.abdisc} how these results imply the
presence of a stronger relationship between luminosity and period than
found in field objects.

\subsection{X-ray to optical flux ratio}
\label{sect.fxfopt}

\subsubsection{Comparison Between Source Classes}

A useful, distance--independent diagnostic for stars detected in both the
optical and X-ray wavelength bands is the ratio of the X-ray flux to the
optical flux.  Figures \ref{fig.fxfopt-ver}a and \ref{fig.fxfopt-ver}b show
plots of this ratio vs \mv\ for nova-like CVs and DNe in the field
(VBR97). The \fx\ values in the 0.5-2.5 keV band from VBR97 were scaled to
be consistent with the slightly absorbed, 1~keV thermal bremmstrahlung
spectra (0.5-2.5 keV) assumed by GHE01a, and used here. The $F_{opt}$
values were derived from $F_{opt}=10^{-0.4V-5.43}$.  In cases where the
distances to the field CVs of VBR97 were unknown, the systems were plotted
on the left side of Figures \ref{fig.fxfopt-ver}a and
\ref{fig.fxfopt-ver}b.

The majority of field DNe have \fxfopt\ $> 0.01$ with the smallest ratios
being dominated by Z Cam systems (with relatively high accretion
rates). The remaining DN classes (U~Gem and SU~UMa systems) mostly have
\fxfopt\ $> 0.1$. Among the nova-like CVs the UX UMa systems have \fxfopt\
values or limits mostly around 0.01.  These ranges show a clear trend that
higher \fxfopt\ values correspond to lower accretion rates for non-magnetic
CVs, as pointed out by VBR97 for this sample of field CVs, and by Richman
(1996). Most magnetic CVs have \fxfopt\ $>0.1$, although it is unclear if
\fxfopt\ is anti-correlated with accretion rate for these systems.

Fig.~\ref{fig.fxfopt}a shows the \fxfopt\ values for the 47~Tuc CV
candidates and MSPs (plus other blue IDs) and Fig.~\ref{fig.fxfopt}b shows
the \fxfopt\ values for the active binary candidates. Limits are shown
where appropriate.  Note the clear differences between Figures
\ref{fig.fxfopt}a and \ref{fig.fxfopt}b, with generally much larger
\fxfopt\ values for the CVs and the MSPs.  With the exception of AKO~9, the
\fxfopt\ values for 47~Tuc CVs and CV candidates are all $>0.1$, suggesting
that the population of 47~Tuc CVs, if dominated by non-magnetic systems,
are low accretion rate CVs (U~Gem and SU~UMa systems, where the latter
sub-class are believed to have the lowest accretion rates among field CVs).
Besides V3, the likely DQ~Her system V1 (GHE01a) has the highest \fxfopt\
value.

Cumulative distributions of \fxfopt\ for the 47~Tuc CVs, active binaries
and MSPs are compared in Fig.~ \ref{fig.cumul-fxfopt}a. Only the objects
detected in $V$ have been included rather than those with limits, with the
exception of 47~Tuc~T.  The clear differences between the CV and the active
binary distributions (and between the MSP and active binary distributions)
are confirmed by the median values given in Table~\ref{tab.fxfopt-median}
and the large KS-test probabilities shown in Table~\ref{tab.fxfopt-ks}. The
MSP distribution only contains 4 systems and the other binary MSPs detected
in the radio fall below our optical detection threshold (but are mostly
detected by \cha), so the true MSP distribution will extend to higher
\fxfopt\ values. The only region of significant overlap between the active
binaries and either the CVs or MSPs is found with 0.1 $<$ \fxfopt\ $<$
0.5. All of these apparently extreme active binaries are found in the faint
group and have radial distributions that are less centrally concentrated
than either the CVs, the MSPs, or the bright active binaries.

\subsubsection{Cataclysmic variables}

In Fig.~\ref{fig.cumul-fxfopt}b the 47~Tuc \fxfopt\ distribution for the
CVs is compared with those found for field CVs using the data given in
VBR97. Table~\ref{tab.fxfopt-median} shows the median values for each
distribution and Table~\ref{tab.fxfopt-ks} compares the KS-test
probabilities. The trend of increasing \fxfopt\ for decreasing accretion
rate (Richman 1996 \& VBR97) is clear, with the UX~UMa (nova-like) systems
and the SU~UMa (DNe) systems at opposite ends of the range, and the Z~Cam
and U~Gem systems falling in between. The 47~Tuc \fxfopt\ values are higher
on average than all of the various CV subgroups, including 3 old novae and
3 double-degenerate systems not shown in Tables \ref{tab.fxfopt-median} and
\ref{tab.fxfopt-ks}.  In a separate calculation we have also included the
field systems with X-ray limits (rather than just detections) given by
VBR97 and this only increases the differences between the \fxfopt\ values
of the 47~Tuc CVs and the field CVs.

To further investigate the origin of differences in \fxfopt\ between the
47~Tuc CVs and the field CVs, we plot in Fig.~\ref{fig.lxmv}a the
cumulative \lx\ distributions of the 47~Tuc CVs and field CVs and a similar
plot in Fig.~\ref{fig.lxmv}b for \mv. Limits in \lx\ (corresponding to
X-ray detection limits, not distance limits) have been included to increase
the sample sizes of the field CVs, since only a subset of the CVs have
estimated distances.  The luminosities for most of the field CV classes are
indistinguishable from each other, with only the DQ~Her systems having a
significantly different (more luminous) distribution, as pointed out by
VBR97. The \lx\ distribution of the 47~Tuc CVs is distinguishable from the
AM~Her, DQ~Her, UX~UMa, Z~Cam, U~Gem and SU~UMa systems at the 97.6\%,
22.4\%, 97.3\%, 72.2\%, 99.3\% and 99.8\% levels respectively, so that only
the DQ~Her systems are obviously a good match to the 47~Tuc CVs. This would
argue against the identification of the 47~Tuc CVs with low accretion rate
CVs like DNe, if the accretion rate is proportional to \lx\ (for example
Warner (1995) suggests $L_{acc} \simeq 50$\lx). However, see the discussion
of the \mv\ distribution below.

Note that the majority of field AM~Her and DQ~Her systems from VBR97 were
X-ray selected (as were the 47~Tuc CVs), unlike other classes of CV,
implying that the relatively high X-ray luminosities could be partially
selection effects, perhaps weakening the classification of the 47~Tuc CVs
with magnetic systems.  For example, crowding in X-rays is likely to have
prevented the detection of some systems with low X-ray luminosities, and
the true cluster distribution will therefore extend to lower
luminosities. Robust limits on incompleteness will be presented in future
publications. Incompleteness will also apply with the field samples,
especially the magnetic systems. For example, although only 2 out of 10 of
the DQ~Her systems in the field sample analysed here involve upper limits,
a significant number of faint DQ~Hers (and other CV classes) may have been
missed in field surveys. The four most luminous DQ~Her systems are at
relatively large distances ($>$400 pc), suggesting that such bright systems
are relatively rare, as pointed out by VBR97. However, despite this
incompleteness at low luminosities, the similarity between the 47~Tuc and
DQ~Her distributions at high luminosities (above $\sim7-8\times10^{31}$
\ergs) is noteworthy, and V1, the brightest CV in 47~Tuc, has already been
shown to be a likely DQ~Her system.

The \mv\ distribution of the 47~Tuc CVs is distinguishable from the AM~Her,
DQ~Her, UX~UMa, Z~Cam, U~Gem and SU~UMa systems at the 8.6\%, 99.97\%,
99.99\%, 98.80\%, 1.03\% and 96.2\% levels respectively. Here, only the
AM~Her and the U~Gem systems are good matches for the 47~Tuc CVs.  A
comparison of Figures \ref{fig.fxfopt-ver} and \ref{fig.fxfopt} shows that
our optical detection limit of $V\sim 23$ (because of crowding; see
\S~\ref{sect.summid}) is a significant source of incompleteness for faint
DNe searches, and may prevent the detection of a significant fraction of
SU~UMa systems. Clearly, the 47~Tuc CVs are significantly fainter in the
optical than the nova-like CVs detected in the field, including the DQ~Her
systems. Since metal-poor secondaries are expected to be brighter than
solar metallicity secondaries, at fixed mass, the difference in accretion
luminosity between the 47~Tuc CVs and field DQ~Hers will be even more
dramatic than suggested by Fig.~\ref{fig.lxmv}b.  Finally, distances are
available for only 2 old novae and none are available for double degenerate
systems (both of these classes of CV may be extremely faint and beyond our
detection limits).

The faint \mv\ values for the 47~Tuc CVs suggest that they have low
accretion rates, consistent with the period/\mv\ analysis, but conflicting
with the interpretation based on their X-ray luminosities.  In fact, an
examination of Fig.~\ref{fig.lxmv} shows that the 47~Tuc CVs have combined
\lx\ and \mv\ distributions that are not consistent with any known class of
field CV (see \S~\ref{sect.cvdisc} for discussion of this issue).

\subsection{qLMXB search}
\label{sect.qlmxb}

As expected, the optical ID for X5/W58 (Edmonds et al. 2002a), a qLMXB, has
a relatively high \fxfopt\ value of 47.8 (Fig.~\ref{fig.fxfopt}; see also
Heinke et al. 2003). Interestingly, the CV candidate V3 has a very similar
\fxfopt\ value. This ratio is higher than that of all known field and
47~Tuc CVs, representing evidence for V3 also being a qLMXB. The
possibility that V3 is a binary containing a neutron star has already been
discussed by GHE01a who argue that V3 may be an enshrouded MSP. For
comparison, source B, the possible qLMXB in NGC 6652 has \fxfopt\ $\sim$9.0
when the $F_{opt}$ value given by Heinke, Edmonds, \& Grindlay (2001) is
appropriately converted, while limits for X7 and the qLMXB in NGC~6397 are
$>$166 (Heinke et al. 2003) and $>$64 respectively. However, we caution
that the \cha\ and GO-8267 data were not obtained simultaneously, and since
V3 is highly variable both optically (see Paper~I) and in X-rays (Verbunt
\& Hasinger 1998; GHE01a) the high \fxfopt\ value derived here could be
explained by a low optical state during the GO-8267 observations, or a high
state during the \cha\ observations.

The two bright (\lx\ $\sim 10^{33}$\ergs) sources and likely qLMXBs in
47~Tuc (X5 and X7) and the likely qLMXBs in NGC~6397 and NGC~6440 are all
soft sources, using the X-ray color definition of Grindlay et al. (2002),
namely xcolor=2.5 log$([0.5-1.5$keV$]/[1.5-6$keV$])$. Therefore, we have
searched for evidence that any of the fainter, soft X-ray sources in 47~Tuc
may be qLMXBs (besides V3). The source AKO~9 is a very soft source but is
more likely a CV than a qLMXB because it has been observed to have at least
2 outbursts without the detection by ROSAT or RXTE of a $10^{36}$
erg~s$^{-1}$ source in 47~Tuc (associated with an LMXB in outburst). The
only other reasonable candidate is W17, a 103 ct source that is a factor of
3 fainter than the qLMXB in NGC~6397.  This source is outside the GO-7503
FoV and unfortunately lies in the middle of a large saturation trail in the
GO-8267 $V$ and $I$ images.  However, useful limits can be set from the
deep and relatively clean $U$ image. The nearest detectable stars to the
X-ray position are at 2.58, 2.91 and 3.04 $\sigma$, and these have $U = $
22.4, 18.3, \& 21.1. The fainter two of these are comparable to the CVs
W44\opt\ and W45\opt\ and the brighter one to a system intermediate between
AKO~9 and V1. We set an upper limit to the detection of a star at the
nominal W17 X-ray position (ie assuming zero error in transforming to \hst\
coordinates) of $U\sim24$. This is suggestive of a faint optical
counterpart like X5 (Edmonds et al. 2002a) or a faint upper limit as with
the qLMXB in NGC~6397 (GHE01b).  Other possibilities, besides a qLMXB, are
that the source is a soft CV (e.g. an AM~Her system) or an active
binary. However, for the latter possibility this source would be a factor
of $\sim$3 brighter than any of the soft (and non-flaring) active
binaries. Followup observations with ACS should be useful in searching for
optical counterparts and determining the nature of this source.

\subsection{Blue Variables} 
\label{sect.bluevar}

As noted in Paper~I, 1V36 may be a faint X-ray source (\lx\ $<
10^{30}$\ergs), arguing in favor of this object being a CV (although
Edmonds et al. 2001 and Edmonds et al 2002b both give examples of faint
X-ray sources with blue, variable optical counterparts that are MSPs, not
CVs). Much deeper \cha\ observations are required (and have been taken) to
confirm this marginal detection of an X-ray source, but there are a number
of arguments against the CV identification: (1) the period of 19.1 hr
(assuming the observed variations are ellipsoidal) is too long (see
Fig.~\ref{fig.permv}).  Ellipsoidal variations should be seen in this low
noise time series if the system is a CV and therefore Roche-lobe filling
(unless the inclination is very low); (2) the position of this star in the
color-color plot is different from most known field and cluster CVs; (3)
the upper limit on \fxfopt\ for 1V36 is lower than that of all 47~Tuc CVs
(see Fig.~\ref{fig.fxfopt}); (4) there is no suggestion of the large
flickering present in other optically bright CVs such as V1, V2 and
W25\opt. Note that (2) and (4) likely rule out the possibility that several
UX~UMa systems with very small \fxfopt\ values might have slipped beneath
our X-ray detection limit.  We therefore believe that the comprehensive
data available for 1V36 does not support the claim by Knigge et al. (2002)
that 1V36 is a strong CV candidate and look forward to seeing what STIS/FUV
spectroscopy and variability says about this system.

Other explanations for 1V36 fare little better.  The variable appears to be
too faint and blue in $V-I$ to be a cluster blue straggler, and the star is
much too bright to be a horizontal branch or blue straggler star in the
SMC. An explanation for this object remains elusive.

All of the other blue variables are uncrowded in the \cha\ image and
consistent with non-detection, and their \hst\ time series show no evidence
for flickering. As with 1V36 this argues against them being CVs.  Here, we
examine possible alternative explanations for these blue variables: (a)
background variable stars from the SMC (e.g. RR Lyraes), (b) detached WD
binaries, (c) very low accretion rate CVs, like those discussed in Townsley
\& Bildsten (2002), and (d) exotic collision products. A combination of
these may also apply.

Considering possibility (a), we plot two SMC RR Lyraes (labeled as R1 and
R2) in the WF3 CMDs of Paper~I. These have periods of 15.2 hr and 8.7 hr
and were independently discovered by two of the authors (Edmonds and
Gilliland) using a ground-based variability study of a 47~Tuc field off the
core (14.5$' \times 14.5'$; for background see Sills et al. 2000). The RR
Lyrae light curves have a distinctive asymmetrical appearance that is
clearly different from that of the blue variables. Also, only 3V07 lies
reasonably close to the expected position of the horizontal branch
instability strip. A second possibility is that the variables are pulsating
blue stragglers in the SMC. These generally have shorter periods and
smaller amplitudes than RR Lyraes, for example the range of periods for the
main pulsation modes of the 6 known SX Phe stars in 47~Tuc are 0.7--2.4 hr
and the range of amplitudes is 0.006-0.085 mags (Gilliland et al. 1998).
So, while the amplitudes overlap, most of the blue variables have periods
that are too long. Also, only 3V07 appears to lie reasonably close to the
expected blue straggler sequence. Consistent with these negative
conclusions, we note that the number of blue SMC stars above the MSTO in
our field is likely to be very small judging by the detection of only 4
such stars in the study of a WFPC2 47~Tuc field by Zoccali et
al. (2001). Proper motion studies with separate F300W epochs may help
confirm the blue variables as 47~Tuc members.

Possibility (b) is attractive because Fig.~\ref{fig.permv}b shows that
several of the blue variables have periods that are reasonably close to,
but longer than, the values given by equation~\ref{eqt.warner} for Roche
lobe filling. However, only 2V30 has colors that are obviously consistent
with those of a MS star/carbon-oxygen (CO) WD binary, and the long period
for this system of 1.18 or 2.36 days is difficult to understand as either
ellipsoidal variations or heating effects.  The colors of 2V08 are very
similar to those of a bright WD, so any MS companion is likely to be
extremely faint. The other objects with all three colors available (1V36,
3V07, 4V05 and the possible ID for W71), have colors, compared to the MS,
that are bluer in $V-I$ than in $U-V$, behavior that is difficult to
reconcile with CO WD-MS star binaries.  However, note from the color-color
CMD given in Paper~I that 3V07 lies very close to the He~WD companion to
the MSP 47~Tuc~U, suggesting that 3V07 may also be a He~WD.  This does not
obviously have an MSP companion, because of the lack of detected X-ray
emission, but it could have a WD companion.

Case (c) appears to be an unlikely explanation because the stars are too
bright to be consistent with the Townsley \& Bildsten (2002) models, and
their colors are not obviously consistent with CO WDs.  Without having a
satisfactory explanation for most of these stars, we turn to possibility (d)
that at least some of them represent an exotic collision product formed
near the center of 47~Tuc. Mathieu et al. (2002) point out, in their
discussion of red stragglers in M67, that stellar encounters are common in
star clusters and that it would not be surprising to discover products of
such encounters (especially binaries) that run counter to standard
evolutionary theory of single stars. Such expectations apply even more so
to 47~Tuc.

\section{Discussion}
\label{sect.disc}

\subsection{Cataclysmic variables}
\label{sect.cvdisc}

We have discovered (or confirmed) optical counterparts for 22 \cha\ sources
that are CV candidates as summarized in Tables~3 and 4 of Paper~I.  We have
included V3 in this list, but, as noted above, it may be a qLMXB. Of these
22 CV candidates, definite variability of some type has been seen for all
of them except W33\opt, W45\opt, W49\opt, W70\opt, W82\opt\ and
W85\opt. The excellent astrometric match between W45 and W45\opt, the tiny
chance of this being a chance coincidence (0.012\%) and the relatively
bright, hard nature of the X-ray source makes this ID secure. The other
candidates have larger possibilities of being chance coincidences (0.8\%,
1.0\%, 1.4\%, 2.3\% and 0.2\%) and so without having independent
information such as variability or an \hal\ excess there remains the
possibility that one or even more of these 5 IDs is not real. Also, W70,
W82 and W85 are relatively soft soures (Fig.~\ref{fig.xcol}) and therefore
one or two of them may be MSPs (see \S~\ref{sect.alter}).  Among the
marginal candidates we believe that W140\opt\ and W55\opt\ have the
greatest chances of being CVs, as explained in Paper~I.

The photometric properties of these stars were summarized in Paper~I.  The
other properties of these systems are summarized as: (1) the \fxfopt\
values are higher, on average, than those of all classes of CV in the field
sample of VBR97, partly because of higher than average X-ray luminosities
and partly because of faint optical magnitudes; (2) the periods, where
available, are generally consistent with expectations for Roche lobe
filling objects; (3) the radial distributions are consistent with those of
the cluster MSPs and the brightest MS-MS binaries, and therefore with WD-MS
star binaries; (4) three of the CVs are eclipsing (W8, W15 and AKO~9),
several show significant flickering (V1, V2, V3, \& W25\opt), two of them
have previously been seen in outburst (V2 and AKO~9) and several others
show large amplitude variations (e.g. W51\opt, V3, W56\opt) that may also
be signs of outbursts.

The period/\mv\ data presented in \S~\ref{sect.permv} and the faint
optical magnitudes presented in \S~\ref{sect.fxfopt}, plus the apparent
lack of non-magnetic novalike systems suggest that the 47~Tuc CVs are
dominated by low accretion rate CVs, i.e. DNe\footnote{From
\S~\ref{sect.qlmxb}, it is unlikely that qLMXBs, with their large \fxfopt\
values, are significantly biasing our results, since based on X-ray
spectral information and \fxfopt\ data, only V3 is a reasonable qLMXB
candidate}. This result is not necessarily inconsistent with the lack of
outbursts seen for 47~Tuc CVs, since some DNe (such as SU~UMa systems) can
have quite long recurrence times (Warner 1995).  Shara et al. (1996)
mention this possibility for 47~Tuc. However, the outbursts seen for V2 and
AKO~9, and the possible outbursts shown in Paper~I (the GO-7503
variability) do not appear to be as dramatic as those seen in SU~UMa
systems, and the \lx\ distributions of the 47~Tuc CVs and field SU~UMas
appear to be incompatible. Also, all of the 47~Tuc CVs except for W21\opt\
are found above the period gap, while nearly all known SU~UMa systems are
found below the period gap. Therefore, we believe it is unlikely that a
significant fraction of the 47~Tuc CVs are SU~UMa systems.

Selection effects in the field will almost certainly conspire against the
detection of low accretion rate CVs (Warner 1995). For example, nova-like
systems are relaitvely bright and blue in the optical. Conversely, SU~UMa
systems like WZ~Sge are extremely faint at optical and X-ray wavelengths,
and have very long recurrence times for outbursts. Also, CVs found in the
field have a range of non-uniform selection criteria.  Although CVs with
very low accretion rates (like those modeled in Townsley \& Bildsten 2002)
will be difficult to detect in 47~Tuc, the sample of cluster CVs is likely
to be relatively complete down to X-ray luminosities of
3--5$\times10^{30}$\ergs. Therefore, given (1) the depth of our
observations, (2) the uniform search methods used, and (3) that the 47~Tuc
CVs are effectively all at the same distance, it may be possible that our
47~Tuc CVs are a more representative sample of CVs than the field objects
of VBR97, and that they therefore show an expected bias towards lower
accretion rate systems.

Is the lower metallicity of 47~Tuc compared to those found in field
(Population~I) CVs likely to result in lower accretion rates? Stehle, Kolb,
\& Ritter (1997) have studied the long-term evolution of CVs with low
metallicity secondaries (Z=$10^{-4}$) and have shown that such systems have
a smaller period gap and a slightly higher mass transfer rate than CVs
where the secondary has a solar composition. This trend therefore does not
explain the low apparent accretion rates for the 47~Tuc CVs. Population
synthesis modeling by Stehle, Kolb, \& Ritter (1997) shows that most
Population~II CVs should be old enough to have already evolved beyond the
period minimum, and will be extremely faint. However, this age effect does
not explain the low accretion rates for the 47~Tuc CVs, since the periods
for the cluster CVs, where known, are nearly all above the period gap (and
most of the CVs with unknown periods should also be found above the period
gap).

A potential problem with the low accretion rate interpretation is that no
known class of field CV has both \lx\ and \mv\ distributions that are
similar to those of the 47~Tuc CVs.  As noted above, the X-ray luminosities
of the 47~Tuc CVs are much higher than those of the U~Gem systems, and are
consistent with those of DQ~Her systems.  So, while the faint optical
magnitudes imply that they have relatively low accretion rates, the high
X-ray luminosities might imply that they have relatively high accretion
rates.  Only one of these possibilities, at most, can be correct.  Since
\mv\ is clearly a better indicator of accretion rate than \lx\ for
non-magnetic CVs (VBR97), and \mv\ is also known to strongly depend on
accretion rate for magnetic systems (e.g. Hessman, G{\" a}nsicke, \& Mattei
2000 and Yi 1994), we speculate that \lx\ for the 47~Tuc CVs is a poor
guide to their accretion rates whether they are magnetic or non-magnetic
systems.  We note that two systems with presumably relatively low accretion
rates are V2 (a DN) and W56\opt, a possible DN. These are two of the
brightest CVs in X-rays, with \lx$ = 6.3 \times 10^{31}$\ergs\ and $1.5
\times 10^{32}$\ergs\ respectively, hinting that high X-ray luminosities
are compatible with low accretion rates.  We encourage theoretical work on
globular cluster CVs to try to explain these results, and caution that the
low accretion rate interpretation given here is not secure until the
combined X-ray and optical luminosities are understood.

Given the relatively high X-ray luminosities of the brightest CVs in 47~Tuc
and the observation that about 25\% of field CVs are magnetic
(Wickramasinghe \& Ferrario 2000), is it possible that most of the brighter
CVs are DQ~Her systems?  Solid evidence exists for DQ~Her-like behavior in
V1 (GHE01a), and the strong resemblance between the spectra of AKO~9 and
GK~Per suggests that AKO~9 is also a DQ~Her system (Knigge et
al. 2001). Unfortunately, these two systems represent just a small fraction
of the total CV population in 47~Tuc, and the period vs \mv\ and \fxfopt\
data presented in \S~\ref{sect.permv} and \S~\ref{sect.fxfopt} shows that
the 47~Tuc CVs are statistically different from DQ~Her systems found in the
field (field DQ~Hers have, on average, significantly brighter optical
magnitudes; Fig.~\ref{fig.lxmv}b).  This contrasts with the similarity
between the \lx\ distributions of the 47~Tuc CVs and the field DQ~Her
systems (see Fig.~\ref{fig.lxmv}a). 

The \lx\ and \mv\ values for the 47~Tuc CVs are consistent with the data
presented for NGC~6397 by Cool et al. (1998), Edmonds et al. (1999) and
GHE01a. The \lx\ values of the NGC~6397 CVs are high (like those in
47~Tuc), since 4 out of 9 systems have \lx\ $> 10^{31.9}$ \ergs
(GHE01b). Optically the NGC~6397 CVs are faint, despite being expected to
have even brighter secondaries than 47~Tuc (because of the low metallicity,
[Fe/H]$ = -1.95$; Harris 1996). The absolute magnitudes of the 4 relatively
bright CVs from Cool et al. (1998) and Edmonds et al. (1999) range between
5.95 and 8.81.

Further information about the NGC~6397 CVs are available because good
quality optical spectra have been obtained. Arguments have been made that
the CVs in NGC~6397 may be dominated by DQ~Hers (Grindlay et al. 1995 and
Edmonds et al. 1999), based on the observation that 3 of the 4 NGC~6397 CVs
with measured optical spectra have relatively strong He~II 4686\AA\ lines,
like those found in field DQ~Hers.  Figure~8 of Edmonds et al. (1999) shows
an analysis of the absolute magnitudes of the accretion disks [\mv(disk)]
for DQ~Her systems and the ratio between the continuum levels of H$\beta$
and H$\alpha$ [C(H$\beta$)/C(H$\alpha$)] for different classes of CV. For
field DQ~Her systems, linear relationships are found between \mv (disk) and
the He~II 4686\AA\ to H$\beta$ equivalent width ratio (HeII/H$\beta$), and
a similar relationship is found between [C(H$\beta$)/C(H$\alpha$)] and
HeII/H$\beta$ (brighter and bluer disks, equivalent to higher accretion
rates, correlate with larger HeII/H$\beta$ values).  The 4 NGC~6397 CVs
discussed in Edmonds et al. (1999) have \mv (disk),
[C(H$\beta$)/C(H$\alpha$)] and HeII/H$\beta$ values that are consistent
with the field DQ~Her systems, but intriguingly, the 3 NGC~6397 CVs with
moderate HeII/H$\beta$ values are all found in the low accretion rate part
of the figure, with low \mv (disk) and low C(H$\beta$)/C(H$\alpha$)
values. Therefore, the low accretion rates apparently found for the 47~Tuc
CVs, combined with the hints of magnetic behavior in a couple of systems,
may be consistent with the NGC~6397 results. Optical spectra of the fainter
CVs in NGC~6397 and the large population of CVs in 47~Tuc are needed to
test whether they also have relatively strong He~II 4686\AA\ (useful
spectra will be difficult to obtain because the CVs are mostly faint and
crowded). Searches for rapid WD spin in the optical and X-rays is also
needed.  Perhaps these observations will identify a new class of low
accretion rate, magnetic CVs in globular clusters.

One final possibility worth considering is that somehow the different
formation mechanism for globular cluster CVs compared to field CVs is
responsible for differences in their respective behavior. To test this
hypothesis we note that CVs in open clusters are much more likely to be
formed from primordial binaries than the 47~Tuc CVs. Interestingly, the two
CVs discovered in the open cluster NGC~6791 by Kaluzny et al. (1997) are
both much bluer (in quiescence) in the $V$ vs $V-I$ CMD than most of the
47~Tuc and NGC~6397 CVs (see Edmonds et al. 1999), suggesting brighter
accretion disks and higher accretion rates than the 47~Tuc and NGC~6397
systems. Indeed, one of the NGC~6791 CVs appears to be either a UX~UMa or
Z~Cam system, with a relatively high accretion rate. Only one other CV is
known in an open cluster, a faint AM~Her system (EU Cancri) in M67
(Gilliland et al. 1991, see Fig.~\ref{fig.permv}; Belloni, Verbunt, \&
Mathieu 1998).  Clearly, any two randomly selected CVs from 47~Tuc would
not look like the NGC~6791 CVs, but obviously the statistics are very poor.
Larger samples of CVs in rich open clusters and low density globular
clusters are needed to see if this effect is significant, or just a
statistical anomaly.

\subsection{Active binaries: comparison with field studies}
\label{sect.abdisc}

We have a total of 29 likely active binaries in 47~Tuc. Twenty seven of
these show statistically significant, mostly periodic, variability, and
most of them are found on, or slightly above the MS or subgiant ridge-line
except for a handful of red stragglers or red straggler candidates. The
total sample of active binaries in 47~Tuc above our X-ray detection
threshold will inevitably be larger, since a sensitive variability study
was only possible with the GO-8267 data.

A key question is whether the active binaries presented here have similar
properties to those of BY~Dras and RS~CVns found in the field.  Here, we
present a brief comparison with the ROSAT All-Sky Survey results of Dempsey
et al. (1997). Figure 4 of Dempsey et al. (1997) shows that the X-ray
luminosities of RS~CVns and BY~Dras are quite different from each other,
with median X-ray luminosities of $\sim1.6\times10^{29}$erg~s$^{-1}$ for
the BY~Dras and $\sim4\times10^{30}$erg~s$^{-1}$ for the RS~CVns and
maximum luminosities of $\sim2.5\times10^{30}$erg~s$^{-1}$ for the BY~Dras
and $\sim1\times10^{32}$erg~s$^{-1}$ for the RS~CVns. These X-ray
luminosities are the quiescent values, as Dempsey et al. (1997) observed
flares on all of their targets during their survey and removed data points
with enhanced count levels.

The red stragglers (W3\opt, W43\opt\ and W72\opt), the red straggler
candidates (W4\opt, W14\opt\ and W38\opt) and the stars near the MSTO like
W9\opt\ and W75\opt\ generally have X-ray luminosities (see GHE01a) that
are in good agreement with the Dempsey et al. (1997) data if the primary
stars are subgiants and these systems are RS~CVns. The two red stragglers
(and possible sub-subgiants) in M67 both have \lx =
$7.3\times10^{30}$erg~s$^{-1}$ (Belloni, Verbunt, \& Mathieu 1998), in good
agreement with the X-ray luminosities of W3\opt\ and W72\opt.  Also, the
relatively bright $V$ mags (and in some cases obviously red colors) of the
X-ray bright BY~Dras (Fig.~\ref{fig.radall}d) are consistent with the
hypothesis that a reasonable number of these objects have at least
partially evolved primaries and hence are RS~CVns.

Among the active binaries with optically fainter IDs the source W47 is
probably returning from a flare, and X-ray variability in other sources
such as W18, W41 and W94 may also be signs of flares, explaining the
relatively high luminosities even though these systems are likely
BY~Dras. The luminosities for the optically faint and apparently non-X-ray
variable systems like W26, W59 and W66 are obviously more extreme. Some of
them could have X-ray variability that is intrinsically quite large, but is
not detected because of the faintness of the sources.  Another possibility
is that they are simply outliers in a large population of objects having a
distribution that is similar to those found in the field. AGB01 point out
that large numbers of BY~Dras may have been missed in their study due to
incompleteness, since the binary frequency determined from the BY~Dra
sample (with periods that are mostly a few days long) is a factor of 17
smaller than that calculated from the shorter period eclipsing and contact
binaries.  Assuming that the binary frequency is the same among the BY~Dras
in the AGB01 study as it is among the short period eclipsing and contact
binaries, and scaling from the sample of 55 AGB01 binaries with $V > 19$
this would imply a population of about a thousand BY~Dras in 47~Tuc.  Since
Figure 4 of Dempsey et al. (1997) shows that $\sim15$\% of the field
BY~Dras have \lx\ $>10^{30}$erg~s$^{-1}$, we would therefore have a deficit
of X-ray detected BY~Dras with \lx\ $>10^{30}$erg~s$^{-1}$, rather than a
surplus. Possible reasons for this deficit are that a reasonable fraction
of the binaries with periods longer than about 1--2 days have been
destroyed by interactions, or that there is an anti--correlation between
\lx\ and period (see below).

The plot of X-ray color vs $V$ in Fig.~\ref{fig.xcol} shows that most of
the optically faint active binaries are relatively soft sources, consistent
with the expected soft X-ray spectra for these objects in quiescence
(Dempsey et al. 1993 and Dempsey et al. 1997).  The source W47 is a
relatively hard source and is variable, and flares are expected to harden
the X-ray spectrum (as noted in Paper~I W64 may show long-term
variability).

If the bright active binaries have a substantial fraction of RS~CVns and
the faint systems are all BY~Dras then there may be an anti-correlation
between \lx\ and $V$ (ie the optically fainter active binaries being
associated with fainter X-ray sources).  We found no statistically
significant evidence for such an anti-correlation, either with or without
the X-ray variables (by contrast the CVs show weak anti-correlations
between \lx\ and $V$, especially when AKO~9 is removed from the
sample). With a much larger number of (mostly undetected) BY~Dras than
RS~CVns it is much more likely that a few very luminous BY~Dras are
detected rather than a few very luminous RS~CVns, possibly explaining the
lack of correlation noted above.

In agreement with the trends shown in Fig.~\ref{fig.rad-bright-faint}d,
there is a significant anti-correlation between period and $V$ mag for the
active binaries. The linear correlations between period and $V$ are $-0.38$
and $-0.48$ when using two different tests (the `kendl1' and `pearsn'
subroutines) from Numerical Recipes (Press et al. 1992), with chance
probabilities of only 0.4\% and 1\%. Presumably the main cause of this
anti-correlation is that the brighter active binaries contain a number of
systems with evolved stars and relatively long periods. However, why have
only the short period BY~Dras (the faint active binaries) been detected in
X-rays? According to Dempsey et al. (1997), there is a fairly weak
correlation between X-ray flux (\fx) as measured at the Earth and rotation
period ($P_{rot}$) of: \fx$\sim P_{rot}^{-0.16\pm0.26}$ for BY~Dras (a
stronger correlation exists for RS~CVns). Assuming that \lx$=
P_{rot}^{-0.16}$ for the active binaries, then the median period of 0.4
days for the faint active binaries (the BY~Dras) and 1.56 days for the
faint AGB01 binaries means that \lx\ would be only 17\% lower for the AGB01
binaries. This is probably not a large enough difference to explain the
very different period distributions, although a steeper slope (but still
remaining within the 1-$\sigma$ error limit) could easily produce a 30-40\%
difference between the X-ray luminosity of the faint active binaries and
the faint AGB01 binaries. We also note that the period distribution of the
Dempsey et al. (1997) sample is very different from ours, since only 3/29
of the Dempsey et al. (1997) BY~Dras with known periods have periods $<$ 1
day, but 15/27 of the 47~Tuc active binaries with known periods have
periods $<$ 1 day (our photometrically selected sample is biased towards
short periods, explaining part of this effect).  With this different sample
of periods (with a much stronger bias towards short periods) it is not
surprising that we find a different \fx/period relationship.

In an important sense our sample of active binaries in 47~Tuc (at fixed
metallicity and age) is significantly cleaner than the Dempsey et
al. (1997) sample containing an inhomogeneous sample of stars with a range
of ages and metallicities. Therefore, the examination of trends such as the
dependence of \fx\ on period may be optimally performed with cluster
samples (we defer this study to a future publication following analysis of
deeper \cha\ observations obtained in late 2002).

\subsection{Are some of the active binaries MSPs?}
\label{sect.alter}

Here, we examine the possibility that some of the active binary candidates
could be MSPs.  Such a population would be very interesting for dynamical
reasons (see \S~\ref{sect.intro}), and it has a direct impact on the number
of MSPs in 47~Tuc estimated from the \cha\ data. The large range in the
estimated number of 47~Tuc MSPs (35--90) given by Grindlay et al. (2002) is
dominated by assumptions about the nature of the active binary population
(see also \S~\ref{sect.revised}).

We begin by noting that the eclipsing binary X-ray sources (W12\opt,
W92\opt, W137\opt, \& W182\opt) are unlikely to be MSPs (or CVs) because
two eclipses would not be expected. Also, the X-ray sources in the AGB01
sample classified as W~UMas (W41\opt, W47\opt, \& W163\opt) have periods
and colors that obey the Rucinski relationship for such systems (AGB01),
therefore making the contact binary explanation a more natural one,
although here we do not rule out an MSP explanation.

The radial distributions of the active binaries presented in
\S~\ref{sect.rad} offer powerful constraints on this issue.  There are
striking similarities between the radial distributions of the active
binaries and the AGB01 binaries (with the active binaries removed) in both
the bright and faint samples (see \S~\ref{sect.rad} and
Fig.~\ref{fig.rad-bright-faint}).  The similarity in the bright samples
suggests that the bright active binaries are not dominated by MSPs with
`normal' MS companions, because then mass segregation would mean that these
$\sim$2.15--2.25\mdot\ objects (1.4\mdot\ neutron star + 0.75--0.85\mdot\
MS star) should be much more concentrated towards the center of the cluster
than the lower mass objects (the radio detected MSPs and the CVs). There is
a chance that some of the bright active binaries could be MSPs with low
mass ($\lesssim 0.2$\mdot) MS star companions that have been heated by
X-rays so that they appear optically like MSTO stars or red stragglers, as
may be occurring for 6397-A (Ferraro et al. 2001). In principle the radial
distribution of these objects could then mimic those of MSTO binaries,
however, this would require a coincidence and would not explain an apparent
absence of MSP binaries with MS star companions having masses close to the
turn-off value of $\sim$0.85\mdot. A second possibility is that
binary-binary and binary-single star interactions resulting in dynamical
kicks may cause the average MSP-MS star binary to be further away from the
center of the cluster than expected for mass segregation of such
objects. However, this would require another coincidental similarity
between radial distributions. It would also be inconsistent with the
difference noted in \S~\ref{sect.rad} between the radial distributions of
the bright and the faint active binaries. This difference probably would
not have been seen if MSPs dominate both the faint and the bright groups
because then the percentage difference in mass between the two groups would
have been much smaller (as shown in Fig.~\ref{fig.rad-bright-faint} the
faint active binaries are unlikely to contain a significant number of MSPs
based on their radial distribution).

When these radial distribution results are combined with the evidence that
(1) the faint active binaries fall above the MS ridge-line just like the
AGB01 sample of binaries (see Paper~I), and (2) that the X-ray luminosities
are consistent with RS~CVn and BY~Dra systems (\S~\ref{sect.abdisc}), we
conclude that the active binaries are indeed dominated by MS-MS binaries.

Is this finding consistent with the properties of the known MSP population?
Of the 16 MSPs with timing positions (Freire et al. 2001) only half of them
are binaries and of these only 5 have companion masses likely to be above
the minimum value (0.085\mdot) for being on the MS (the other 3 have masses
of $\sim 0.02-0.03$\mdot). It is likely that these 5 MSPs {\it all} have He
WD companions based on the radio data (Camilo et al. 2000), e.g.  47~Tuc~U
has an \hst-identified He~WD companion (Edmonds et al. 2001) while a faint,
blue star and possible ID has been found for 47~Tuc~T and faint limits can
be set in the optical for 47~Tuc~H. Only one MS companion was found
(47~Tuc~W; Edmonds et al. 2002b) in the full sample of 20 MSPs from Camilo
et al. (2000). Therefore, if the sample of 16 MSPs is representative of the
possible much larger sample of MSPs (Camilo et al. 2000) then optical
companions are rare and MS companions even more so.  If a large population
of MSPs with MS companions does exist, and we are detecting many of them in
X-rays, then the MSPs would have to be intrinsically radio-faint to avoid
any of them being detected by Parkes.  However, the relatively flat
correlation between X-ray and radio luminosity shown by Grindlay et
al. (2002) implies that the faintest MSPs in the radio are less faint as
X-ray sources.


Given the small number of detections of cluster MSPs with likely
non-degenerate companions (two; 47~Tuc~W and 6397-A) it is not yet possible
to conclude whether such MSPs are generally relatively bright in X-rays but
faint in the radio.  The MSP 6397-A would have been easily detected by
Parkes if it were in 47~Tuc, given the radio luminosity of 5 mJy kpc$^2$
quoted by D'Amico et al. (2001). The radio luminosities of the 47~Tuc MSPs
range between 0.81 and 10.9 mJy kpc$^2$, with 12 out of 14 of them having
radio luminosities less than that of 6397-A. The MSP 47~Tuc~W, however, is
one of the faintest detected MSPs, and W34, if it is an MSP, is below the
detection limit of Camilo et al. (2000). The X-ray luminosities for 6397-A
and 47~Tuc~W are slightly higher than the most luminous `normal' MSP in
47~Tuc.

Without having a deeper sample of MSPs, and detailed radial velocity
information for the active binaries it is very difficult to rule out the
possibility that any {\it individual}, non-eclipsing, optical variable has
an MSP companion, especially in view of the cases of W29\opt\ and 6397-A
(and possibly also W34\opt).  The former object is very crowded optically
and was only discovered using difference image analysis (Edmonds et
al. 2002b), but was found on close study to have blue colors. Therefore,
W29\opt\ would have been classified as a CV instead of an active binary if
not for the remarkable period and phase coincidence with 47~Tuc~W
discovered by Edmonds et al. (2002b), perhaps implying that one or two of
the faint CV candidates could be an MSP. None of the active binary
candidates show blue colors, and they also show no hint of the large
amplitude variability seen in either W29\opt\ or W34\opt, despite in some
cases having periods that are not much longer than those of W29\opt\
(suggesting similar levels of variability caused by irradiation
from an MSP, if present).

One potential source of MSPs masquerading as active binaries are the red
stragglers. For example, the companion to the MSP 6397-A was previously
identified as a BY~Dra by Taylor et al. (2001) on the basis of H$\alpha$
emission, a CMD position lying above the MS and optical variability (this
star is clearly a red straggler). However, it was subsequently shown by
Ferraro et al. (2001) to have an MSP companion based on an accurate radio
timing position for the MSP.  Naturally, it is possible for a binary system
to be both an active binary and to have an MSP companion, since secondaries
in short period binaries like 6397-A are undoubtedly tidally locked and
rapidly rotating, and hence may have substantial chromospheric activity (as
suggested for 6397-A by Orosz \& van Kerkwijk 2002).

Although the 47~Tuc red stragglers and 6397-A have similar optical
properties, there are some key differences between their X-ray properties.
While 6397-A is a reasonably hard source (as is 47~Tuc~W), the two red
stragglers W3 and W72 both have very soft X-ray colors and the two red
straggler candidates W4 and W14 also have soft colors (see
Fig.~\ref{fig.xcol} and Fig.~\ref{fig.xcmd}). Of the red stragglers, only
W43 (in GO-7503) has a hard color.  Generally, if most of the active binary
candidates are MSPs and are like 47~Tuc~W and 6397-A, then they should
mostly be hard sources. Yet, of the GO-8267 active binaries with X-ray
color determinations from Grindlay et al. (2002) only 4 (not including
W47\opt) are hard while 9 (not including W4) are soft.

The leading candidates for MSPs hidden within our active binary sample
include W43, a faint, moderately hard, variable X-ray source (like 6397-A)
with an optical ID that falls in a similar region of the optical CMD to
6397-A, as noted above. No clear optical variability is present, with our
limited quality time series. We note that while W43 is clearly harder than
the other red stragglers, this could be because of flaring activity,
explaining the variability. Both W23 and W64 are moderately hard sources
that were not observed to vary during the \cha\ observation of GHE01a, and
are brighter than all of the other active binaries (except W47). However,
as explained earlier, W64 may show long-term variability.

The leading candidates for identifications of other 47~Tuc~U-like MSPs (not
yet detected in the radio), are the weak, soft, non-variable X-ray sources,
with faint, blue, optical companions that do not show long-term
variability. These are W82 and W85 (see Fig.~\ref{fig.xcol} and
Fig.~\ref{fig.xcmd}), with plausible optical IDs (lying outside the GO-8267
FoV) without obvious long-term variability. They are beyond the faint limit
of the GO-7503 data for short-term variability studies. The ID for W70
shows no evidence for periodic variation in our time series and only
marginal evidence for non-periodic variations, but we do not yet have
long-term variability information for this object. These sources could be
CVs, since AM~Her systems for example are known to be soft X-ray sources
(VBR97).  Our high quality H$\alpha$ observations with \hst/ACS may help
distinguish between these two possibilities, since He~WDs should show broad
H$\alpha$ absorption lines, and CVs strong H$\alpha$ emission lines, as
observed for the CVs in NGC~6397 (Grindlay et al. 1995 \& Cool et al. 1998)
and NGC~6752 (Bailyn et al. 1996).

\subsection{Revised estimate of total number of MSPs}
\label{sect.revised}

Finally, we examine the soft, faint, apparently non-variable X-ray sources
that have no plausible (optically variable or photometrically unusual)
counterparts (see Fig.~\ref{fig.xcmd}b). Grindlay et al. (2002) listed the
sources in this category, noting in particular the subset of these sources
that fall in the GO-8267 FoV: W4, W5, W6, W24, W28, \& W98 (we now add the
source W71 to this list). As noted by Grindlay et al. (2002) these sources
include the leading candidates to be MSPs. By correcting for the incomplete
spatial coverage of the WFPC2 GO-8267 FoV, and for incompleteness in the
detection of radio-detected MSPs as soft X-ray sources, Grindlay et
al. (2002) estimated the presence of at least 19 MSPs with \lx\ $\gtrsim
10^{30}$\ergs\ that do not have radio counterparts with timing positions,
in addition to the 16 MSPs with such positions. This estimate obviously
increases if some of the active binaries identified with soft sources are
instead MSPs, but as argued above, few are expected. Also, the estimate by
Grindlay et al. (2002) implicitly assumes that 2/9 of the MSPs have X-ray
colors $< 1.0$, so a small fraction of misidentified `active binaries' with
harder X-ray colors (like W9 or W59) that are really MSPs, do not change
these statistics.

Based on a study of the optical data, is it reasonable that the 7 soft
sources just discussed are all MSPs, rather than CVs or active binaries in
some cases? First, as already noted, W4 is a candidate red straggler and
W24 may be an RS~CVn, and both of them have higher X-ray count levels than
the brightest detected MSP (except for the hard X-ray source
47~Tuc~W). Also, W98 is located close to several giant stars so it may also
be an RS~CVn.  To estimate how many active binaries, or perhaps CVs, may
have missed detection because of crowding, we have examined the stellar
count levels present in the GO-8267 data around the positions of the 7 soft
sources. Figure \ref{fig.testsky} shows the mean stellar counts in an
annulus (between 0\farcs9 and 1\farcs35) around each: (1) CV, (2) bright
active binary, (3) faint active binary (4) MSP, (5) unidentified X-ray
source, and (6) candidate MSP (the 7 sources discussed above). Optical
positions are used for (1), (2), (3) and (6), radio positions for (4) and
X-ray positions for (5).  The median count level for the candidate MSPs is
higher than any of these other groups, and is 2.0 times larger than the
median count level for the MSPs (the two distributions differ at the 98.8\%
level using the KS-test). This suggests that some of the 7 soft sources are
active binaries (or CVs) that have been missed because of crowding. We
tentatively classify W4, W24 and W98 as active binaries, based on their
promixity to bright stars, and classify the 4 remaining sources as
MSPs. This implies a population of 13 extra MSPs for a total of 29 MSPs
with \lx\ $> 10^{30}$ \ergs.

\subsection{Summary of optical identifications}
\label{sect.summid}

Incorporating the optical identifications reported here and in Paper~I
(plus radio/MSP identifications from GHE01a and Grindlay et al. 2002) and
just considering the GO-8267 FoV alone (where our high quality time series
coverage is spatially complete) a significant fraction of the 78 sources
are active binaries (34.6\%) compared to 19.2\% for CVs and 9.0\% for MSPs,
with the source W46/X7 being a qLMXB (1.3\%). These sources, where X-ray
colors are available, are labeled in the X-ray CMD shown in
Fig.~\ref{fig.xcmd}.  The remaining 27 (34.6\%) sources in the GO-8267 FoV
have uncertain identifications, or have no plausible ID.  Of these sources,
we use the X-ray luminosity and color, or the plausible match-up with
bright, blue or marginally variable stars to tentatively identify W10, W16,
W17, W20, W32, W35, W37 and W140 as CVs, and W4, W24, W37, W71, W93, W141
and W168 as active binaries leaving 13 faint sources that we identify as
possible MSPs (see Table~\ref{tab.undet8267}). This leaves us with final
fractions of 42.3\% active binaries, 29.5\% CVs and 26.9\% MSPs (plus the
single qLMXB), although some of the 13 weak X-ray sources could be CVs or
active binaries, lowering the MSP number.  The CV fraction is very similar
to that calculated by GHE01a, but the active binary fraction is
significantly higher than the 15\% estimated by GHE01a.  Correspondingly,
the MSP estimate, and the one given in \S~\ref{sect.revised}, are smaller
than the value of $\sim$50 MSPs originally presented for the 108 sources in
GHE01a (that estimate was based on incomplete analysis of the optical
data).

\subsection{Conclusion and Prospects}
\label{sect.summ-prosp}

Based on the period/\mv\ analysis and the \mv\ distribution presented here,
the 47~Tuc CVs appear to have lower accretion rates than field CVs systems
found above the period gap.  Theoretical work on CV evolution predicts that
the fainter, shorter period CVs expected to exist in 47~Tuc (but not
observed because of crowding) should have even lower accretion rates than
the bright ones that we do observe (see Di Stefano \& Rappaport 1994 and
Townsley \& Bildsten 2002).  The suggestive evidence for low accretion
rates and DQ~Her type behavior in CVs in NGC~6397, combined with: (1) the
evidence for magnetic behavior in a couple of the 47~Tuc CVs (V1 and AKO~9)
and (2) the high (DQ~Her-like) X-ray luminosities of the 47~Tuc CVs implies
that magnetic activity may play an important role in globular cluster CVs.
However, more work is needed to understand the unusual \mv\ and \lx\
distributions for the CVs in both 47~Tuc and NGC~6397.

Perhaps the most unexpected result, when compared with other clusters, is
that the active binaries outnumber the CVs. In NGC~6397, the detected CVs
outnumber the active binaries by a factor of $\sim$2--3 (based on deeper
\cha\ data than obtained for 47~Tuc), but significant depletion of the MS
binary population in NGC~6397 is likely to have occurred during and after
core collapse. In NGC~6752, the detected CVs outnumber the active binaries
by a factor of $\sim$4--10, but the NGC~6752 \cha\ data is not as deep as
the 47~Tuc data, and several unidentified soft sources may still be
identified with active binaries, given optical data of higher quality.

With this large population of detected active binaries in 47~Tuc, many of
them soft sources, we find support for the lower range given by Grindlay et
al. (2002) for the number of MSPs in 47~Tuc with \lx\ $ > 10^{30}$\ergs\
($\sim$30--40). It is possible that a few of the active binaries may really
be MSPs, and we continue to search for new objects that are similar to
47~Tuc~U or 47~Tuc~W, but, given the spatial, photometric, X-ray spectral
and timing information available for the observed active binaries, a large
population of MSPs masquerading as active binaries is unlikely.  The lower
estimated number of MSPs in 47~Tuc means that the ratio between the number
of MSPs and the number of CVs is smaller, and closer to the corresponding
value found in NGC~6397 by GHE01b.

\begin{figure}
\epsscale{0.7}
\plotone{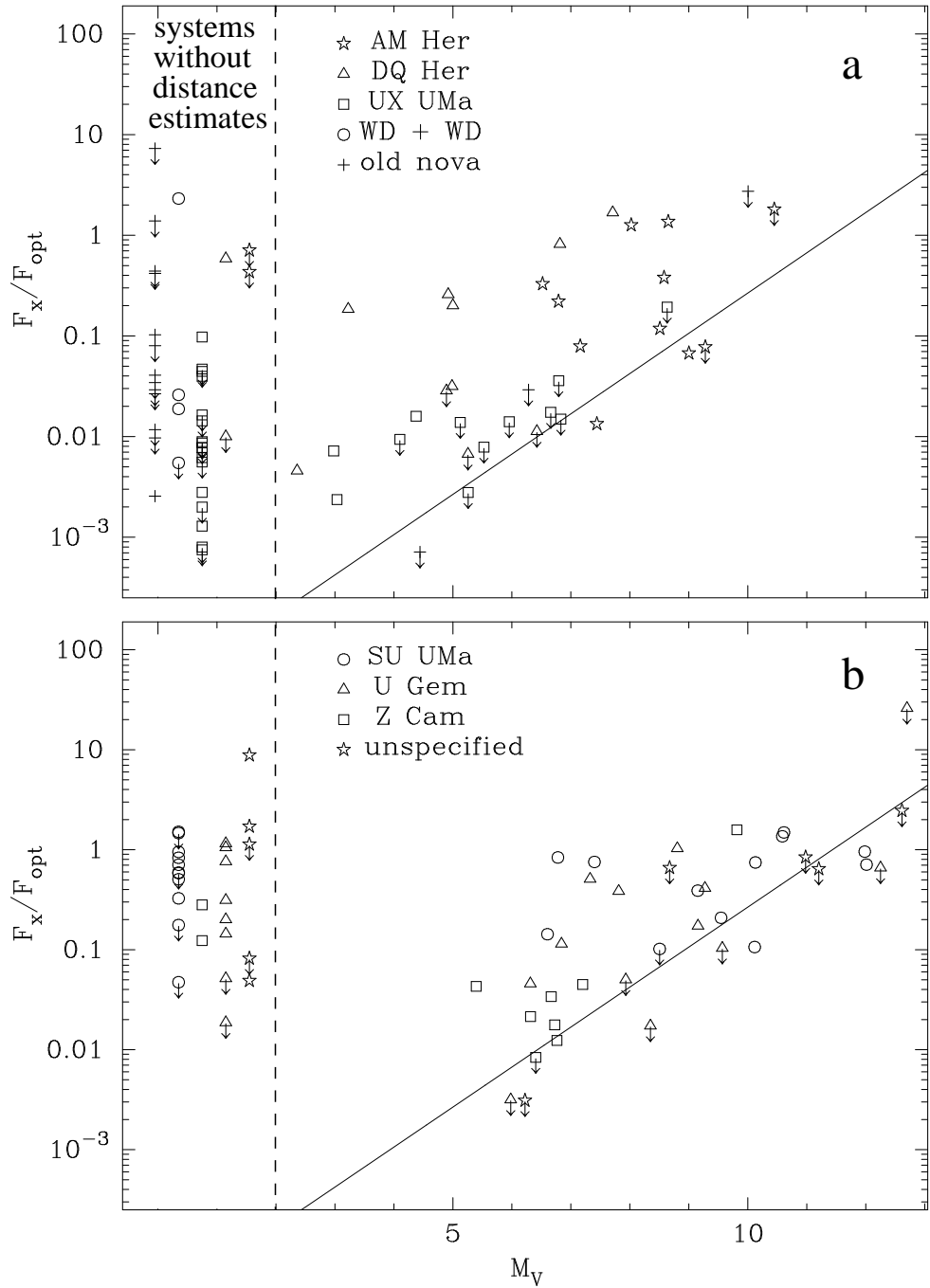}
\caption{Plots of \fxfopt\ vs $M_V$ for the field CVs from
VBR97. Fig.~\ref{fig.fxfopt-ver}a shows the nova-like CVs and
Fig.~\ref{fig.fxfopt-ver}b shows the DNe. Field systems are plotted where
the distance is known, otherwise the systems are shown in the left portion
of the figures. The straight line shows \lx\ $= 10^{30}$\ergs.}
\label{fig.fxfopt-ver}
\end{figure}

\begin{figure}
\plotone{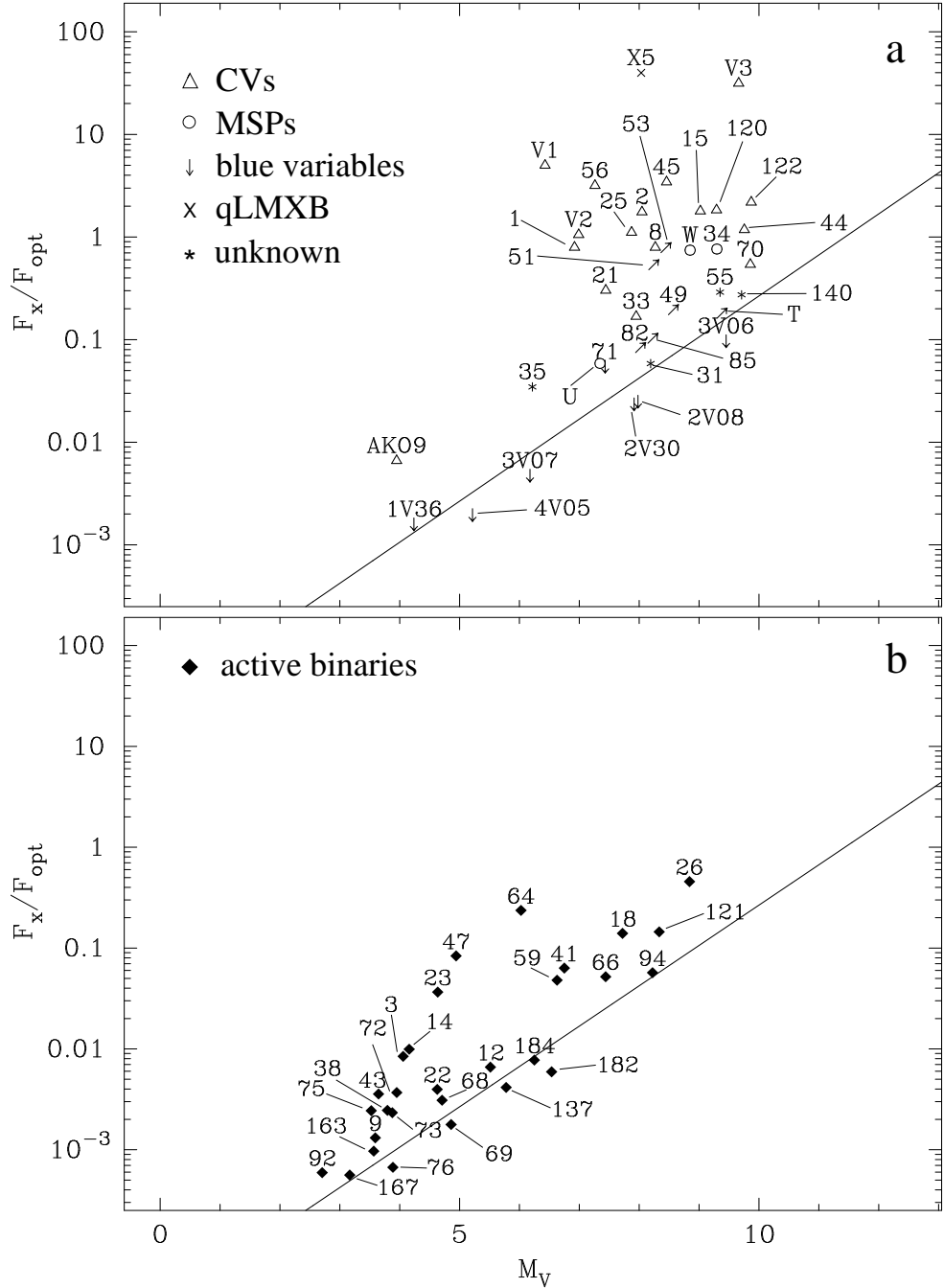}
\caption{Plots of \fxfopt\ vs $M_V$ for the 47~Tuc optical identifications
of \cha\ sources. The symbols used as the same as Fig.~\ref{fig.rms},
except that arrows are shown for limits in \fx\ (blue variables) or $M_V$
(faint CVs in GO-7503 and 47~Tuc~T).}
\label{fig.fxfopt}
\end{figure}

\begin{figure}
\epsscale{0.75}
\plotone{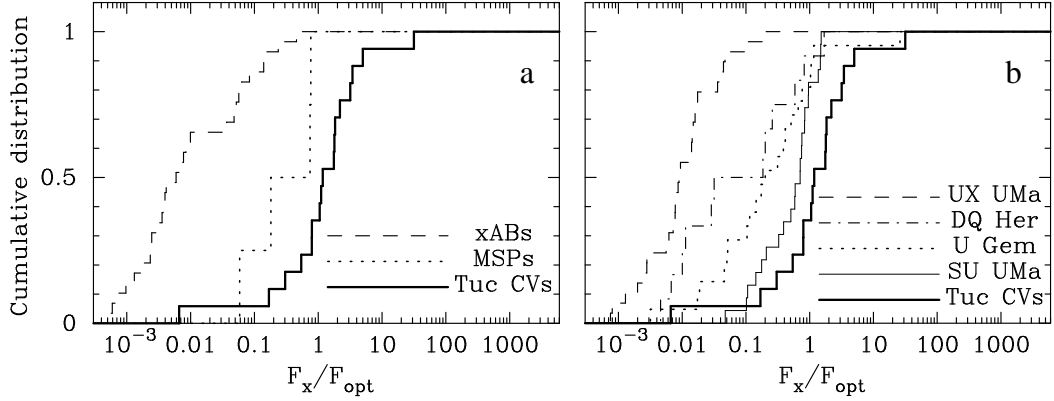}
\caption{Cumulative plots of \fxfopt\ for the 47~Tuc CVs compared to the
47~Tuc active binaries and MSPs (Fig.~\ref{fig.cumul-fxfopt}a) and various
classes of field CV from the work of VBR97
(Fig.~\ref{fig.cumul-fxfopt}b). For the non-magnetic field CVs the
accretion rates decrease from left to right. These results imply that the
non-magnetic CVs in 47~Tuc have low accretion rates, although the \fxfopt\
values are biased by the unusually high \lx\ values for the 47~Tuc CVs.}
\label{fig.cumul-fxfopt}
\end{figure}

\begin{figure}
\epsscale{0.75}
\plotone{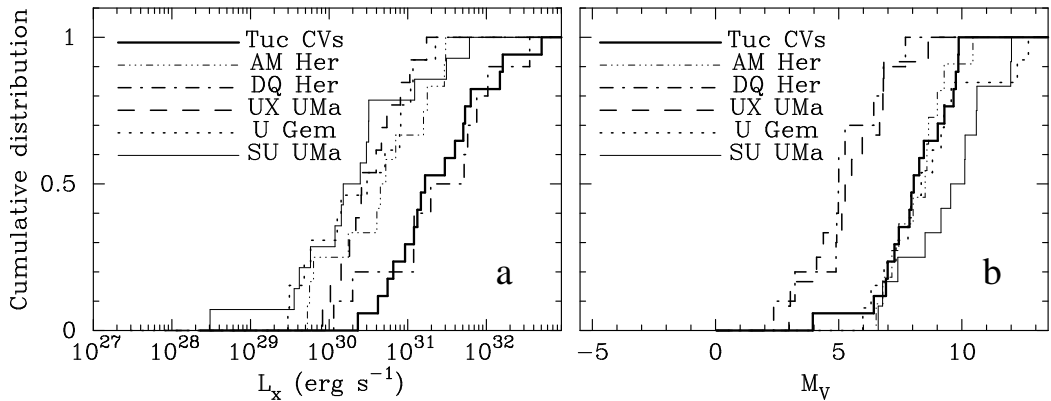}
\caption{Cumulative plots of \lx\ (Fig.~\ref{fig.lxmv}a) and \mv\
(Fig.~\ref{fig.lxmv}b) for the 47~Tuc CVs compared to the field CVs from
VBR97. Upper limits have been included in the field \lx\
distributions to increase the sample size. Note the similarity between the
\lx\ distribution of the 47~Tuc CVs and the field DQ~Her systems, and the
similarity between the \mv\ distribution of the 47~Tuc CVs and the field
U~Gem and AM~Her systems.}
\label{fig.lxmv}
\end{figure}

\begin{figure}
\epsscale{0.7}
\plotone{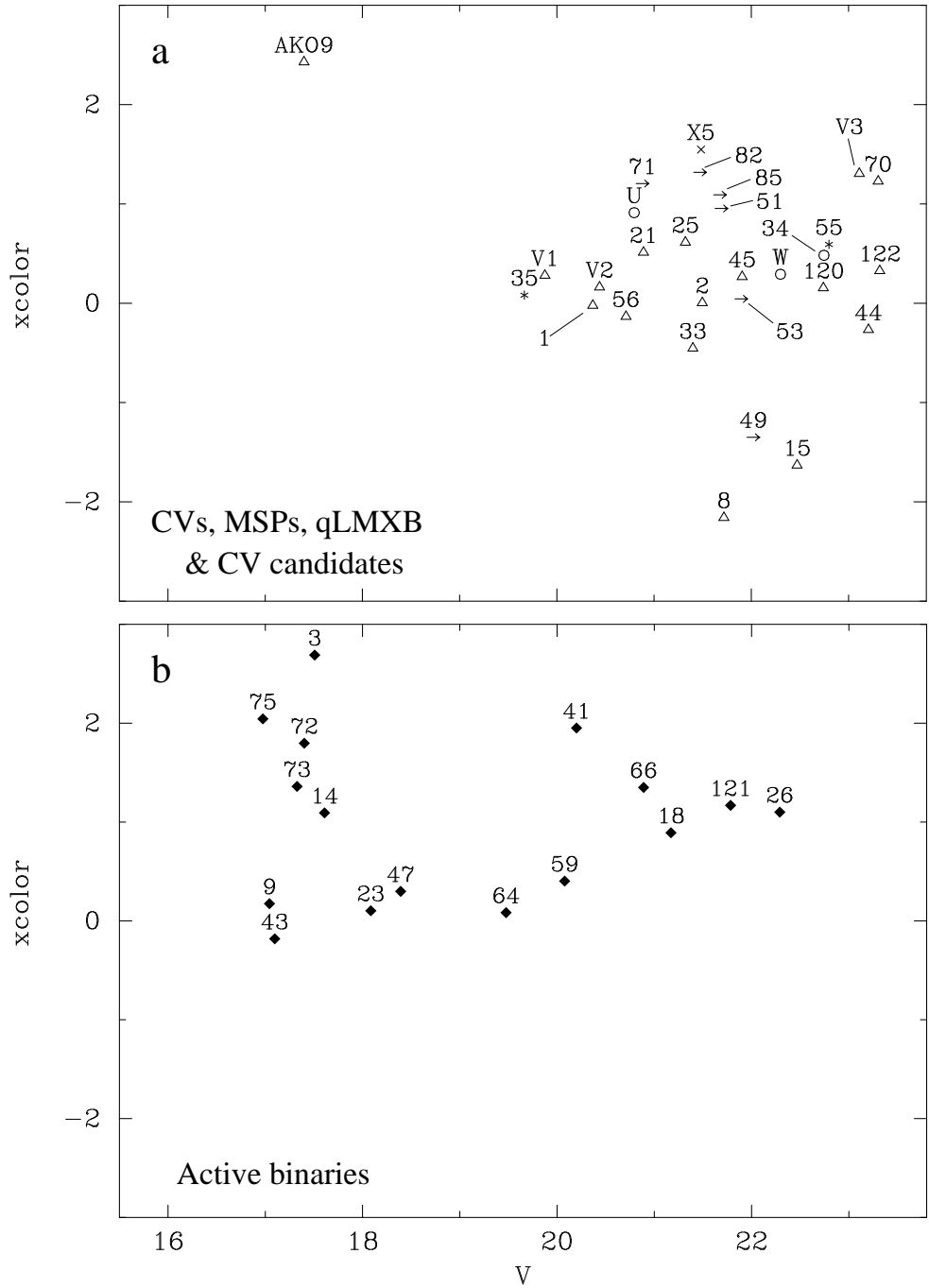}
\caption{X-ray color plotted against $V$. For the X-ray color we use the
definition of GHE01a and the selection criteria of Grindlay et al. (2002)
for faint sources. Fig.~\ref{fig.xcol}a shows the CVs, MSPs, the qLMXB X5
and CV candidates, and Fig.~\ref{fig.xcol}b shows the active binaries.}
\label{fig.xcol}
\end{figure}

\begin{figure}
\epsscale{0.7}
\plotone{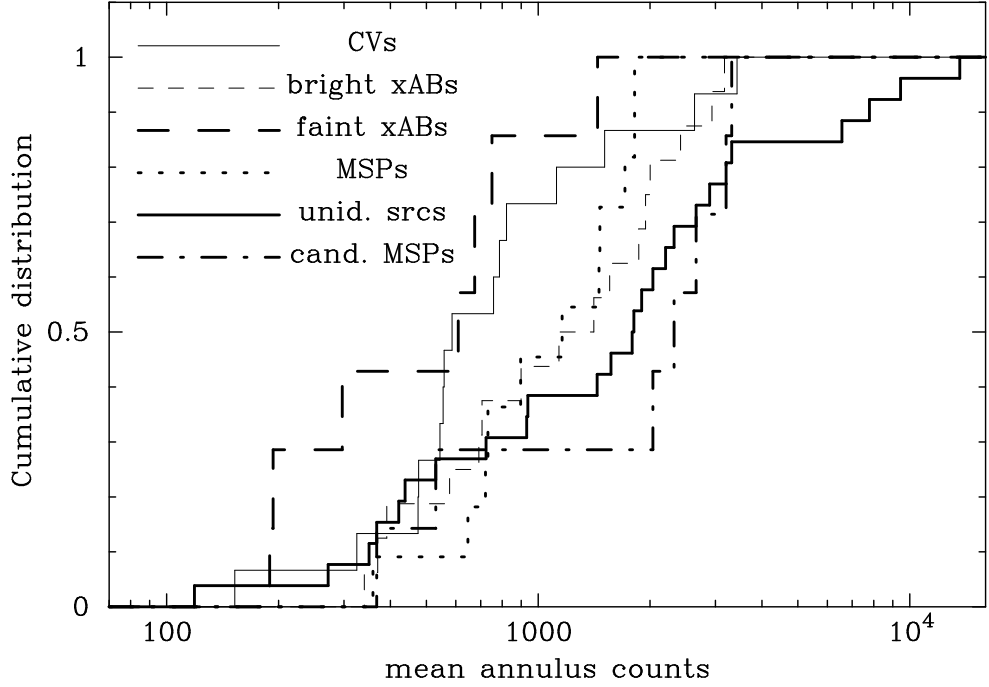}
\caption{Cumulative distribution of mean count values (stellar and
background) for annular regions in the GO-8267 $V$-band image centered on
CVs, bright ($V<19$) active binaries, faint ($V>19$) active binaries, MSPs,
unidentified X-ray sources and candidate MSPs (soft X-ray sources without
optical IDs). The optical positions are used for the CVs and active
binaries, the optically-corrected radio positions are used for the MSPs
(except for 47~Tuc~W where the optical position is used), and the
optically-corrected X-ray positions are used for the unidentified sources
and the candidate MSPs. }
\label{fig.testsky}
\end{figure}

\begin{figure}
\epsscale{1.0}
\plotone{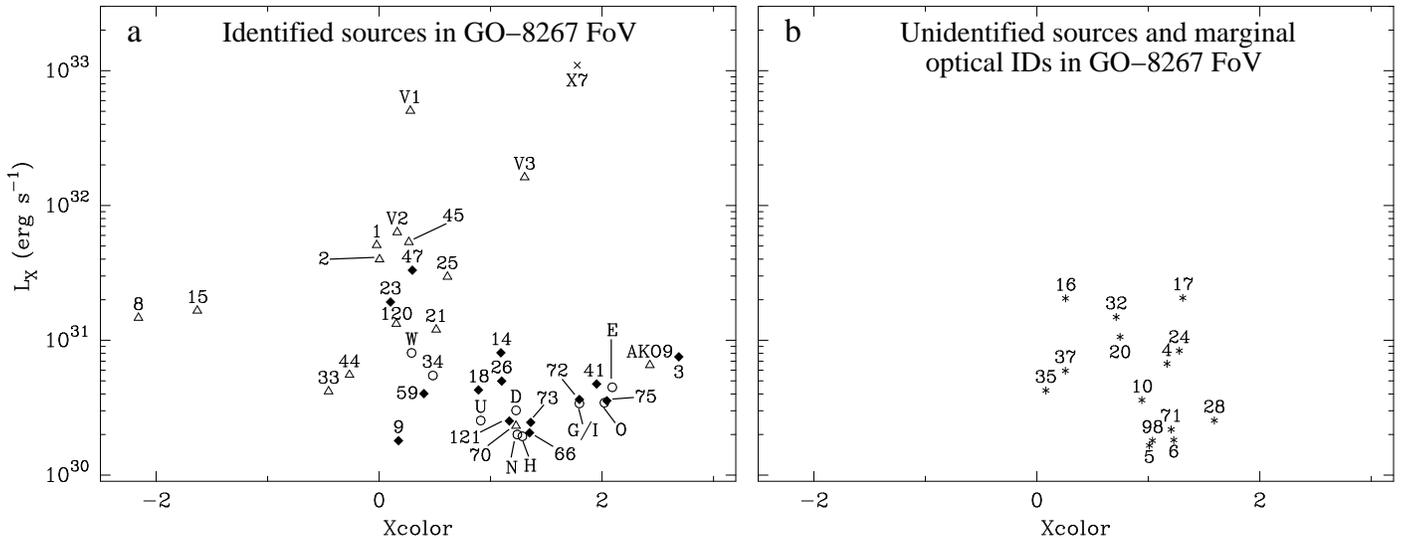}
\caption{X-ray CMD plotting \lx\ in the 0.5-2.5 keV band versus X-ray color
(as defined by GHE01a).  Fig.~\ref{fig.xcmd}a shows the optical
identifications of CVs, MSPs and active binaries in the GO-8267 dataset,
the radio/X-ray identifications of MSPs without optical IDs (47~Tuc~D, E,
G/I, H, N; Camilo et al. 2000; Freire et al. 2001) and X7 (W46) an X-ray
identified qLMXB (GHE01a and Heinke et al. 2003) without an optical ID
(Edmonds et al. 2002a). Fig.~\ref{fig.xcmd}b shows sources with X-ray
colors but lacking optical or radio identifications (some of these have
marginal optical IDs). }
\label{fig.xcmd}
\end{figure}

A comparison of Figures \ref{fig.fxfopt-ver} and \ref{fig.fxfopt}a shows
that we are sensitive to detection of most classes of CV except for the
faintest DNe (SU~UMas), and possibly also double degenerate systems (where
the statistics are poor). There are good prospects for finding at least
some of the fainter CV population not yet detected here by using the much
deeper X-ray limits set by the 300~ks ACIS-S observation.  Although the
gains will be limited near cluster center because of X-ray crowding, the
faintest CVs should have low-mass companions of $\sim 0.1-0.3$\mdot, and
with WD masses of $\sim$0.55\mdot\ the total system masses should be less
than or equal to the MSTO mass. Therefore, many of these objects should be
found in relatively uncrowded regions more than 20-30$''$ away from the
cluster center.  Unfortunately, optical identifications will be difficult
or impossible for many of these objects because of crowding and the high
background levels present in the \hst\ images, a penalty for studying
objects in this massive, concentrated globular cluster.

The 300~ks followup \cha\ observations will provide much better quality
spectral, timing and positional information, especially for the faintest
sources discussed here, and these data should be very useful for
distinguishing between MSPs and active binaries.  Many new, faint sources
will also likely be discovered, dominated by MSPs and active binaries.  New
radio observations are obviously needed to search for the $\sim$200
undetected MSPs believed to exist in 47~Tuc (Camilo et al. 2000).

\acknowledgments

We acknowledge comments on the CVs from Brian Warner, and a number of
extremely helpful comments from the referee, Christian Knigge.  This work
was supported in part by STScI grants GO-8267.01-97A (PDE and RLG) and
HST-AR-09199.01-A (PDE).

\clearpage

\clearpage 

\clearpage

\tabletypesize{\scriptsize}

\begin{deluxetable}{rrl}
\tablecolumns{3}
\tablewidth{0pc}
\tablecaption{\fxfopt\ values for 47~Tuc sources and field CVs}
\tablehead{\colhead{Source type} & \colhead{Number} & \colhead{\fxfopt} \\
\colhead{} & \colhead{} & \colhead{(median)} \\
}
\startdata

   xABs & 29 & 0.0059 \\
   MSPs &  4 & 0.46   \\
    CVs & 17 & 1.18   \\
\cutinhead{Field CVs\tablenotemark{a}}
UX Umas &  9 & 0.0072 \\
 Z Cams &  9 & 0.043  \\
AM Hers &  9 & 0.22   \\
DQ Hers &  8 & 0.23   \\
 U Gems & 13 & 0.39   \\
SU UMas & 18 & 0.73   \\

\tablenotetext{a}{CVs from the study of Verbunt et al. (1997)}

\enddata
\label{tab.fxfopt-median}
\end{deluxetable}


\tabletypesize{\scriptsize}

\begin{deluxetable}{crrrrrrrr}
\tablecolumns{9}
\tablewidth{0pc}
\tablecaption{KS-test probabilities for \fxfopt\ distributions being different}
\tablehead{\colhead{Source} & \colhead{SU UMas} & \colhead{U Gems} & \colhead{DQ Hers} &
\colhead{AM Hers} & \colhead{Z Cams} & \colhead{UX UMas} & \colhead{CVs} & \colhead{MSPs} \\
}
\startdata

 U Gems\tablenotemark{a}& 77.39 &       &       &       &       &        &        & \\
DQ Hers\tablenotemark{a}& 86.13 & 13.40 &       &       &       &        &        & \\
AM Hers\tablenotemark{a}& 97.06 & 42.68 &  0.82 &       &       &        &        & \\
 Z Cams\tablenotemark{a}& 99.84 & 99.13 & 87.52 & 92.22 &       &        &        & \\
UX Umas\tablenotemark{a}&100.00 & 99.99 & 99.26 & 99.65 & 98.13 &        &        & \\
    CVs\tablenotemark{b}& 97.72 & 98.20 & 93.36 & 98.44 & 99.93 & 100.00 &        & \\
   MSPs\tablenotemark{b}& 41.28 &  1.13 &  1.43 &  4.57 & 90.46 &  99.11 &  97.97 & \\
   xABs\tablenotemark{b}&100.00 &100.00 & 99.73 & 99.93 & 99.74 &   8.17 & 100.00 & 99.01 \\

\tablenotetext{a}{CVs from Verbunt et al. (1997)}
\tablenotetext{b}{Optically identified sources in 47~Tuc}

\enddata
\label{tab.fxfopt-ks}
\end{deluxetable}


\tabletypesize{\scriptsize}

\begin{deluxetable}{rccc}
\tablecolumns{4}
\tablewidth{0pc}
\tablecaption{\cha\ sources with marginal or no \hst\ counterparts in the
GO-8267 data}
\tablehead{\colhead{Source} & \colhead{Chip \#} & \colhead{Crowding level} 
  & \colhead{Plausible}\\
\colhead{W\#} & \colhead{} & \colhead{} 
  & \colhead{ID\tablenotemark{a}}\\}
\startdata

 20 & 1 & A & CV? \\
 24 & 1 & B & AB? \\
 28 & 1 & A & MSP? \\
 31 & 1 & B & MSP? \\
 32 & 1 & D & CV? \\
 35 & 1 & C & CV? \\
 37 & 1 & B & CV? \\
 39 & 1 & B & 47~Tuc~O\tablenotemark{b} \\
 46 & 1 & B & qLMXB \\
 96 & 1 & A & MSP?\tablenotemark{c} \\
 97 & 1 & C & MSP?\tablenotemark{c} \\
 98 & 1 & D & AB? \\
141 & 1 & C & AB?\tablenotemark{c} \\
  4 & 2 & D & AB? \\
  5 & 2 & D & MSP? \\
 10 & 2 & D & CV? \\
 13 & 2 & C & 47~Tuc~N\tablenotemark{b} \\
 16 & 2 & D & CV? \\
 17 & 2 & D & CV? \\
 19 & 2 & C & 47~Tuc~G,I\tablenotemark{b} \\
 67 & 2 & C & 47~Tuc~D\tablenotemark{b} \\
 74 & 2 & C & 47~Tuc~H\tablenotemark{b} \\
 91 & 2 & A & MSP?\tablenotemark{c} \\
 95 & 2 & B & MSP?\tablenotemark{c} \\
142 & 2 & D & MSP?\tablenotemark{c} \\
168 & 2 & D & AB?\tablenotemark{c} \\
115 & 3 & C & MSP?\tablenotemark{c} \\
  6 & 4 & D & MSP?\tablenotemark{c} \\
  7 & 4 & D & 47~Tuc~E\tablenotemark{b} \\
 71 & 4 & C & MSP? \\
 93 & 4 & C & AB?\tablenotemark{c} \\
 99 & 4 & A & MSP?\tablenotemark{c} \\
140 & 4 & A & CV?\tablenotemark{c} \\

\tablecomments{Grade for crowding level: (A) Source in uncrowded region; 
(B) Source in moderately crowded region; (C) Source in extremely crowded
region; (D) Source in region affected by saturation}
\tablenotetext{a}{Plausible source identification based on X-ray spectral
information and nearby objects in the optical}
\tablenotetext{b}{MSP from Camilo et al. (2000)}
\tablenotetext{c}{Source does not formally have an X-ray color, using the
 criterion of Grindlay et al. (2002)}

\enddata
\label{tab.undet8267}
\end{deluxetable}


\end{document}